# Information–Theoretic Capacity and Error Exponents of Stationary Point Processes under Random Additive Displacements

Venkat Anantharam and François Baccelli
ananth@eecs.berkeley.edu, Francois.Baccelli@ens.fr

October 11, 2018

## Abstract

This paper studies the Shannon regime for the random displacement of stationary point processes. Let each point of some initial stationary point process in $\mathbb{R}^n$ give rise to one daughter point, the location of which is obtained by adding a random vector to the coordinates of the mother point, with all displacement vectors independently and identically distributed for all points. The decoding problem is then the following one: the whole mother point process is known as well as the coordinates of some daughter point; the displacements are only known through their law; can one find the mother of this daughter point? The Shannon regime is that where the dimension $n$ tends to infinity and where the logarithm of the intensity of the point process is proportional to $n$. We show that this problem exhibits a sharp threshold: if the sum of the proportionality factor and of the differential entropy rate of the noise is positive, then the probability of finding the right mother point tends to 0 with $n$ for all point processes and decoding strategies. If this sum is negative, there exist mother point processes, for instance Poisson, and decoding strategies, for instance maximum likelihood, for which the probability of finding the right mother tends to 1 with $n$. We then use large deviations theory to show that in the latter case, if the entropy spectrum of the noise satisfies a large deviation principle, then the error probability goes exponentially fast to 0 with an exponent that is given in closed form in terms of the rate function of the noise entropy spectrum. This is done for two classes of mother point processes: Poisson and Matérn.

This class of problems is of independent relevance in point process theory, where this Shannon regime has hardly been considered. Its practical interest to information theory comes from the explicit connection that we also establish between this problem and the estimation of error exponents in Shannon's additive noise channel with power constraints on the codewords. Each error exponent for the point process problem leads to an exponent for Shannon's additive noise channel through some simple reparameterization. When specifying this connection in the white Gaussian noise case in the Poisson and Matérn cases, we recover the random and expurgated error exponents of Shannon and Gallager in the regime considered by Poltyrev. More generally, the present paper gives new bounds on error exponents for Shannon's additive noise channel that hold for all stationary and ergodic noises with the above properties and that match the best known bounds in the white Gaussian noise case in the high SNR limit.

# 1 Introduction

This paper is focused on the capacity and on error exponents for an additive noise channel with codebooks consisting of a stationary and ergodic point process of the Euclidean space and with decoding strategies that are stationary relative to the underlying codebook point process.

More general codebooks and decoding rules can of course be considered, but our primary aim in the present paper is to investigate what the machinery of stationary point processes, in particular Palm calculus, and what the simplest parametric classes of stationary point processes, like Poisson or Matérn, can bring to the evaluation of capacity and error exponents.

Since the codewords of such codebooks have no power (norm) constraint, the notion of capacity is not that of Shannon's additive noise channel but that first considered by Poltyrev in the additive white Gaussian noise (AWGN) case in [17]. As we shall see, the connection with Shannon's initial viewpoint (on capacity and error exponents for codebooks subject to a power constraint) is simply obtained by looking at restrictions of the point process to balls with a radius determined by the power constraint. All results (on capacity or error exponents) obtained within the infinite Euclidean space, stationary point process setting of the present paper hence translate to results on Shannon's additive noise channel.

This infinite Euclidean space, stationary point process setting was first introduced in [1], where AWGN error exponents were investigated. The present paper is focused on this point process setting in the case of *general stationary and ergodic noise processes*. In this case, the capacity is determined by the differential entropy rate of the noise process and the associated error exponents are derived from a Large Deviation Principle (LDP) either on the entropy spectrum of the noise process or on its empirical measures. This connection between error exponents and Large Deviation Principles for the last two objects is the main new result of the paper.

This present paper is structured in three parts. The first part (from Section 2 to Section 4) is meant to give a concise summary of the main result. For this, we define notation (Section 2), then the setting (Section 3) and finally the main result (Section 4).

In the second part (from Section 5 to Section 8), we define and analyze the Shannon-Poltyrev capacity and error exponents for general stationary point processes with additive displacement noise. Section 5 focuses on the definition of the Shannon-Poltyrev capacity. The latter is first evaluated in the white Gaussian noise case (Subsection 5.1), then for white noise with a general distribution (Subsection 5.2) and finally for general stationary and ergodic noise (Subsection 5.3). Section 6 gives representations of the error probability within this setting in terms of the Palm probability of the underlying point process, with a special emphasis on the case of Poisson point processes. These representations hold in the general stationary and ergodic noise case (Subsection 6.3) and admit specific incarnations in the white or colored Gaussian cases (Subsections 6.1 and 6.2). These representations are the basis for the analysis of error exponents. Section 7 gives other representations of the error probability based on the mass transport principle and on perturbations analysis. The Shannon-Poltyrev error exponents of a stationary point process are defined and studied in Section 8. The core result bears on the case of a stationary and ergodic noise process and assumes that the entropy spectrum of the latter satisfies an LDP. The error exponents are then deduced from an almost direct application of the Laplace-Varadhan lemma to the Palm probability integral representation of the error probability. In the present paper, we limit ourselves to error exponents associated with Poisson and Matérn point processes and to noise processes with densities. In order to provide a smooth presentation, we follow a general pattern that consists in looking first at the white Gaussian noise (wgn) case, then the white noise (wn) case, and finally at the general stationary and ergodic noise (sen) case. For the latter, our two main examples are the colored Gaussian noise (cgn) and the Markov noise (mn) case.

The third part (from Section 9) is devoted to the connections between the results of the second part and the classical results of Information Theory, namely the Shannon capacity/error exponent theorems for additive noise channels with power constraints. Subsection 9.1 recalls the classical information theoretic viewpoint on the capacity and the error exponent of additive white Gaussian noise channels in the case with restrictions on the power of codewords. The general connection between Poltyrev's capacity and error exponents and Shannon's is the object of Subsection 9.2. We show that, in the stationary ergodic noise case, the Shannon capacity admits an expansion when power tends to infinity, where the second term of the expansion is the Poltyrev capacity. This property is used to relate the Shannon capacity for general noise and the extension of Poltyrev's capacity to general noise which is given in Subsection 5.2. This view-point allows one to retrieve error exponents which achieve Poltyrev's bounds in the AWGN case. It also leads to new bounds on error exponents for the additive noise channel with or without constraints.

## 2 Notation, Basic Definitions

Throughout the paper, all logarithms are to the natural base. We denote by:

- $B^n(x, r)$ the open ball with center $x$ and radius $r$ in the $n$-dimensional Euclidean space $\mathbb{R}^n$;

- $V_B^n(r) = \frac{r^n \pi^{n/2}}{\Gamma(\frac{n}{2}+1)}$; the volume of $B^n(x, r)$;

- $SS^{n-1}(r)$ the sphere of radius $r$ in $\mathbb{R}^n$ centered at the origin;

- $A_{n-1}$ the area of $SS^{n-1}(1)$: $A_{n-1} = nV_B^n(1) = \frac{n\pi^{n/2}}{\Gamma(\frac{n}{2}+1)}$;

- $\text{Cube}^n(a)$ the cube of side length $a$ centered at the origin in $\mathbb{R}^n$;

- $|\mathcal{T}|$ the cardinality of a finite set $\mathcal{T}$.

Throughout the paper, for all basic definitions pertaining to

- point process theory, see [6];
- information theory, see [4];
- large deviations theory, see [7].

## 3 Random Displacement of a Stationary Point Process

### 3.1 Notation

For all integers $n$, let

- $(K_n, \mathcal{K}_n)$ be some measurable space;

- $\mathbb{M}^n(K_n)$ (resp. $\mathbb{M}^n$) denote the set of simple marked point measures (resp. simple point measures) on $\mathbb{R}^n \times K_n$ (resp. $\mathbb{R}^n$);

- $\mathcal{M}^n(K_n)$ (resp. $\mathcal{M}^n$) be the smallest $\sigma$-algebra containing the events $\nu^n(B \times L) = k$ (resp. $\nu^n(B) = k$), $B$ ranging over the Borel sets of $\mathbb{R}^n$, $L$ over the measurable sets of $K_n$ and $k$ over the nonnegative integers (for all questions pertaining to measurability issues, in particular the completion of sigma fields, the reader should refer to [12]);

- $\mathbb{M}_0^n(K_n)$ (resp. $\mathbb{M}_0^n$) be set of all simple marked point measures (resp. simple point measures) with an atom having its first coordinate at 0 (resp. an atom at 0);

- $\mathcal{M}_0^n(K_n)$ (resp. $\mathcal{M}_0^n$) be the trace $\sigma$-algebra of $\mathcal{M}^n(K_n)$ (resp. $\mathcal{M}^n$);

- $(\Omega^n, \mathcal{F}^n, \mathbb{P}^n, \theta_t^n)$ be a probability space endowed with an ergodic and measure preserving shift $\theta_t^n$ indexed by $t \in \mathbb{R}^n$.

Each $\nu \in \mathbb{M}^n$ has a representation of the form

$$\nu = \sum_k \epsilon_{x_k},$$

with $\epsilon_x$ the Dirac measure at $x$ and $\{x_k\}$ the atoms of $\nu$. To all $\nu \in \mathbb{M}^n$, and all atoms $x$ of $\nu$, one associates the Voronoi cell of point $x$ w.r.t. $\nu$ denoted by $\mathcal{V}(x, \nu)$ (see e.g. [16]). Here, we take each $\mathcal{V}(x, \nu)$ an open set.

Each $\mu \in \mathbb{M}^n(K_n)$ has a representation of the form

$$\mu = \sum_k \epsilon_{x_k, m_k},$$

with $\{(x_k, m_k)\}$ the atoms of $\mu$, where $x_k \in \mathbb{R}^n$ and $m_k \in K_n$. The set $\{x_k\}$ is the set of *points* of $\mu$ and the set $\{m_k\}$ is its set of *marks*.

A marked point process $\mu^n$ on $\mathbb{R}^n \times K_n$ is a measurable map from $(\Omega^n, \mathcal{F}^n)$ to $(\mathbb{M}^n(K_n), \mathcal{M}^n(K_n))$ such that for all $t \in \mathbb{R}^n$,

$$\mu^n(\theta_t(\omega)) = \tau_t(\mu^n(\omega)),$$

where $\tau_t(\mu^n)$ is the translation of $\mu^n$ by $-t \in \mathbb{R}^n$. That is, if $\mu^n = \sum_k \epsilon_{T_k^n(\omega), M_k^n(\omega)}$, then

$$\mu^n(\theta_t(\omega)) = \sum_k \epsilon_{-t + T_k^n(\omega), M_k^n(\omega)}.$$

We will use the following notation for such stationary marked point processes:

- $\lambda_n$ is the intensity of $\mu^n$;

- $R_n$ is the normalized logarithmic intensity of $\mu^n$: $\lambda_n = e^{nR_n}$;

- $\mathbb{P}_0^n$ is the Palm probability of $\mu^n$ (by convention, under $\mathbb{P}_0^n$, $T_0^n = 0$);

- $\mathcal{V}_k^n$ is the Voronoi cell of point $T_k^n$ w.r.t. the point process $\mu^n$ i.e. $\mathcal{V}_k^n = \mathcal{V}(T_k^n, \mu^n)$.

## 3.2 Decoding Functional

In what follows, we focus on specific marks of a stationary point process, which are regions of the Euclidean plane associated to its points. These marks are built from a decoding functional $\mathbb{C}^n$, namely a measurable map from $(\mathbb{R}^n \times \mathbb{M}^n, \mathcal{B}(\mathbb{R}^n) \times \mathcal{M}^n)$ to the space of open sets of $\mathbb{R}^n$ (which can be endowed with a topology and can hence be made a measurable space [15]). This map is assumed to be such that if $\nu = \sum_k \epsilon_{x_k^n}$ is a simple point measure (a codebook), then

- For all $k$,
$$\mathbb{C}^n(x, \nu) = O_k^n, \quad \forall x \in \mathcal{V}(x_k^n, \nu),$$
with $O_k^n$ an open set of $\mathbb{R}^n$ called the *decoding region* of $x_k^n$ (the last function is not defined on the boundary of the Voronoi cell - we will not need this);

- The sets $\{O_k^n\}_k$ form a *tessellation of* $\mathbb{R}^n$, namely such the sets $O_k^n$ are all disjoint and the union of their closures is $\mathbb{R}^n$;

- The mapping $\mathbb{C}^n$ is *shift–compatible*, namely such that
$$\mathbb{C}^n(x - t, \tau_t(\nu)) = \mathbb{C}^n(x, \nu) - t,$$
for all $\nu$, $t$ and $x$.

The decoding region of the point $T_k^n$ of the stationary point process $\nu^n$ is defined as
$$\mathcal{C}_k^n = \mathbb{C}^n(T_n^k, \mu^n).$$
It follows from the shift compatibility assumption that $\mathcal{C}_k^n$ is a mark of point $T_k^n$.

Below, to the stationary point process $\nu^n = \sum_k \epsilon_{T_k^n}$ and the decoding functional $\mathbb{C}^n$, we will associate the marked point process $\mu^n = \sum_k \epsilon_{T_k^n, (D_k^n, \mathcal{C}_k^n)}$, where, for all $k$, $D_k^n$ is an independent random vector of $\mathbb{R}^n$, called the *displacement vector* of point $T_k^n$ (the default assumption is that the displacement sequence $\{D_k^n\}_k$ is i.i.d. and independent of the marked point process $\{T_k^n, \mathcal{C}_k^n\}_k$) and $\mathcal{C}_k^n$ is the decoding region of point $T_k^n$.

Each point $T_k^n$ of $\nu^n$ may be thought of as a codeword. When this codeword is transmitted, the channel adds to it the displacement vector $D_k^n$, so that the received signal is $Y_k^n = T_k^n + D_k^n$. The decoding strategy associated with $\mathbb{C}^n$ expects that when $T_k^n$ is transmitted, then the received signal lands in $\mathcal{C}_k^n$. An error happens if this is not the case.

The canonical example to keep in mind, which is motivated by the AWGN channel, is when the vectors associated to the individual points of the process are independent and identically distributed (i.i.d.) zero mean Gaussian random vectors each with i.i.d. coordinates and independent of the points. Then the decoding region of a point is its Voronoi cell in the realization of the point process [16].

The most general setting concerning the noise (or displacement vectors) will feature a real–valued, centered, stationary and ergodic stochastic process $\{\mathcal{D}_l\}_l$ and displacement vectors $D_k^n$ independent of the point process, i.i.d. in $k$, and with a law defined by $D_0^n = (\mathcal{D}_1, \ldots, \mathcal{D}_n)$ for all $n$. As we shall see, more elaborate though natural decoding functionals/regions then show up, determined by the law of $\{\mathcal{D}_l\}_l$. Within this general setting, for all $n$, when $\mu^n$, $\mathcal{C}^n$ and the law of $D^n$ are given, we will define the associated *probability of error* as

$$p_e(n) = \lim_{W \to \infty} \frac{\sum_k \mathbb{1}_{T_n^k \in B^n(0,W)} \mathbb{1}_{Y_k^n \notin \mathcal{C}_k^n}}{\sum_k \mathbb{1}_{T_n^k \in B^n(0,W)}} \qquad (1)$$

and the *probability of success* as $p_s(n) = 1 - p_e(n)$. The limit in (1) exists almost surely and it is non-random. This follows from the assumption that the marked point process $\mu^n$ with marks $(D_k^n, \mathcal{C}_k^n)$ is stationary and ergodic. The pointwise ergodic theorem implies that

$$p_e(n) = \mathbb{P}_0^n \left( Y_0^n \notin \mathcal{C}_0^n \right) = \mathbb{P}_0^n \left( D_0^n \notin \mathcal{C}_0^n \right) . \qquad (2)$$

As we shall see, $p_e(n)$ also has the following equivalent representation:

$$p_e(n) = \mathbb{E}_0^n \left( \sum_{k \neq 0} 1_{Y_k^n \in \mathcal{C}_0^n} \right) = \mathbb{E}_0^n \left( \sum_{k \neq 0} 1_{T_k^n + D_k^n \in \mathcal{C}_0^n} \right) , \qquad (3)$$

(this is proved in Theorem 3.2-sen below).

## 4 Main Results

Consider the general situation where the noise is defined from some stationary and ergodic process $\mathcal{D}$. Assume that:

1. for all $n$, $D^n$ admits a density $f^n$ on $\mathbb{R}^n$ with a well defined and finite differential entropy;

2. $\{\mathcal{D}_k\}_k$ admits a finite differential entropy rate $h(\mathcal{D})$.

For all stationary and ergodic point processes $\mu^n$ of normalized logarithmic intensity $-h(\mathcal{D}) - \ln(\alpha)$ and all jointly stationary decoding regions $\mathcal{C}^n = \{\mathcal{C}_k^n\}_k$, let

$$p_e(n, \mu^n, \mathcal{C}^n, \alpha, \mathcal{D}) \qquad (4)$$

denote the probability of error associated with these data, as defined in (2).

**Capacity.** Then, our main capacity result (Theorem 1-sen and 2-sen below) is that

- For all sequences of point processes with normalized logarithmic intensity $-h(\mathcal{D}) - \ln(\alpha)$, $\alpha < 1$, and for all choices of decoding regions $\mathcal{C}_k^n$,

$$\lim_{n \to \infty} p_e(n, \mu^n, \mathcal{C}^n, \alpha, \mathcal{D}) = 1;$$

- For all $\alpha > 1$, there exists a sequence of point processes, all with normalized logarithmic intensity $-h(\mathcal{D}) - \ln(\alpha)$, and jointly stationary decoding regions $\mathcal{C}_k^n$ such that
$$\lim_{n \to \infty} p_e(n, \mu^n, \mathcal{C}^n, \alpha, \mathcal{D}) = 0.$$

We write $c(\mathcal{D})$ for $-h(\mathcal{D})$ and call it the capacity of the noise process $\mathcal{D}$.

**Error exponents.** Let
$$p_{e,opt}(n, \alpha, \mathcal{D})$$
denote the infimum of $p_e(n, \mu^n, \mathcal{C}^n, \alpha, \mathcal{D})$, where the infimum bears on all stationary and ergodic point processes $\mu^n$ with normalized logarithmic density $-h(\mathcal{D}) - \ln(\alpha)$. Consider the following *error exponent*:
$$\underline{\eta}(\alpha, \mathcal{D}) = \liminf_n -\frac{1}{n} \ln(p_{e,opt}(n, \alpha, \mathcal{D})).$$

Our main result error on this exponent (Theorem 4-sen below) states that if, in addition to Assumptions 1. and 2. above,

3 the entropy spectrum of $D^n$, namely the law of the random variable $-\frac{1}{n} \ln(f^n(D^n))$, satisfies a Large Deviation Principle with a good and convex rate function $I(x)$, then for all $\alpha > 1$,
$$\underline{\eta}(\alpha, \mathcal{D}) \geq \inf_u \{F(u) + I(u)\},$$
where
$$F(u) = \left( \ln(\alpha) + h(\mathcal{D}) - \sup_{s \leq u}(s - I(s)) \right)^+.$$

This result is based on an analogue of random coding, namely on Poisson point processes. An extension of this result to the analogue of expurgated error exponents is also given in terms of hard exclusion Matérn point processes (Theorem 3-sen-Matérn).

**Connections with Shannon's error exponents.** Let $\mathcal{D}$ be some stationary and ergodic noise satisfying the above assumptions and with variance $\sigma^2$. Let $C_P(\mathcal{D})$ denote the Shannon capacity of the additive noise channel with power constraint $P$ on the codewords and with noise $\mathcal{D}$. Let $\underline{\mathcal{E}}(R, P, \mathcal{D})$ denote Shannon's lower error exponent for this channel and for rate $R$ (see Section 9 for precise definitions). Our main result on the connection between Shannon's error exponent and the exponent defined above is given in Theorem 5-sen, which implies that for all $\alpha > 1$,
$$\underline{\mathcal{E}}\left(C_P(\mathcal{D}) - \ln(\alpha) - \frac{1}{2}\ln\left(1 + \frac{\sigma^2}{P}\right), P, \mathcal{D}\right) \geq \underline{\eta}(\alpha-, \mathcal{D}),$$
where
$$\underline{\eta}(\alpha-, \mathcal{D}) = \lim_{\epsilon \to 0} \underline{\eta}(\alpha - \epsilon, \mathcal{D}).$$

# 5 Additive Noise Capacity of a Stationary Point Process

## 5.1 White Gaussian Noise under MLE Decoding

Let $\mu^n$ be a sequence of stationary point processes of intensity $\lambda_n$ in $\mathbb{R}^n$. The displacement vectors are assumed to be independent of the point processes. Each displacement vector is assumed Gaussian, with i.i.d. coordinates having mean zero and variance $\sigma^2$. The displacement vectors are assumed i.i.d. from point to point. Further, we take here $\mathcal{C}_k^n = \mathcal{V}_k^n$, the Voronoi cell of $T_k^n$ w.r.t. the point configuration $\mu^n$. It is easy to see that this corresponds to MLE decoding. The generalization of Theorem 1-wgn to arbitrary decoding functionals as defined in Section 3.2 will be a consequence of Theorem 1-wn in Section 5.2. The main results in this case are:

**Theorem 1-wgn.** *For all stationary point processes $\mu^n$ such that for some subsequence $n_k \to \infty$, $\liminf_{k \to \infty} R_{n_k} > \frac{1}{2} \ln \frac{1}{2\pi e\sigma^2}$, the probability of error defined in (1), for MLE decoding, is such that $\lim_{k \to \infty} p_e(n_k) = 1$.* □

**Theorem 2-wgn.** *Let $\mu^n$ be a Poisson process of intensity $\lambda_n = e^{nR_n}$. For all subsequences $n_k \to \infty$ such that $\limsup_{k \to \infty} R_{n_k} < \frac{1}{2} \ln \frac{1}{2\pi e\sigma^2}$, we have, for MLE decoding, $\lim_{k \to \infty} p_e(n_k) = 0$.* □

**Proof of Theorem 1-wgn.** Consider the ball of radius $\sqrt{n}L_n$ with the same volume as the Voronoi cell of point 0; namely $L_n$ is the r.v. such that

$$V_B^n(\sqrt{n}L_n) = \text{Vol}(\mathcal{V}_0^n).$$

By an immediate monotonicity argument using the shape of the Gaussian density,

$$\mathbb{P}_0^n(D_0^n \notin \mathcal{V}_0^n) \geq \mathbb{P}_0^n\left(D_0^n \notin B^n(0, \sqrt{n}L_n)\right).$$

But

$$\mathbb{P}_0^n\left(D_0^n \notin B^n(0, \sqrt{n}L_n)\right) = \mathbb{P}^n\left(\frac{1}{n}\sum_{i=1}^n D_0^n(i)^2 > L_n^2\right),$$

with $D_k^n(i)$ the $i$-th coordinate of vector $D_k^n$. By the strong law of large numbers (SLLN), for all $\epsilon > 0$,

$$\mathbb{P}_0^n\left(\left|\frac{1}{n}\sum_{i=1}^n D_0^n(i)^2 - \sigma^2\right| \geq \epsilon\right) = \eta_\epsilon(n),$$

with $\eta_\epsilon(n) \to_{n \to \infty} 0$. Hence for all $\epsilon > 0$,

$$\begin{aligned}\mathbb{P}_0^n(Y_0^n \notin \mathcal{V}_0^n) &\geq \mathbb{P}_0^n(\sigma^2 - \epsilon \geq L_n^2) - \eta_\epsilon(n) \\ &= 1 - \mathbb{P}_0^n\left(V_B^n(\sqrt{n(\sigma^2 - \epsilon)}) < \text{Vol}(\mathcal{V}_0^n)\right) - \eta_\epsilon(n).\end{aligned}$$

Now by Markov's inequality,

$$\mathbb{P}_0^n\left(\mathrm{Vol}(\mathcal{V}_0^n) > V_B^n(\sqrt{n(\sigma^2-\epsilon)})\right) \leq \frac{\mathbb{E}_n^0(\mathrm{Vol}(\mathcal{V}_0^n))}{V_B^n(\sqrt{n(\sigma^2-\epsilon)})}$$

and by classical results on the Voronoi tessellation,

$$\mathbb{E}_n^0(\mathrm{Vol}(\mathcal{V}_0^n)) = \frac{1}{\lambda_n} = e^{-nR_n}.$$

Hence

$$p_e(n) \geq 1 - \eta_\epsilon(n) - \frac{e^{-nR_n}}{V_B^n(\sqrt{n(\sigma^2-\epsilon)})},$$

so that, in order to conclude the proof, it is enough to show that the last fraction tends to 0 as $n$ tends to infinity, for some $\epsilon > 0$.

Under the assumptions of the theorem, there is some $c$ with $\sigma > 2c > 0$ such that, for all sufficiently large $k$, with $n = n_k$, we have

$$e^{-nR_n} \leq e^{\frac{n}{2}\ln 2\pi e(\sigma-2c)^2}.$$

In addition, for all $0 < \epsilon < 2c\sigma - c^2$,

$$V_B^n(\sqrt{n(\sigma^2-\epsilon)}) = \frac{(n(\sigma^2-\epsilon))^{n/2}\pi^{n/2}}{\Gamma(\frac{n}{2}+1)} \geq e^{\frac{n}{2}\left(\ln(2\pi e(\sigma-c)^2)\right)},$$

for $n$ large enough. Hence

$$\frac{e^{-nR_n}}{V_B^n(\sqrt{n(\sigma^2-\epsilon)})} \leq \left(\frac{\sigma-2c}{\sigma-c}\right)^n \to_{n\to\infty} 0.$$

$\square$

Theorem 2-wgn follows from the analysis or error exponents (Theorems 4-wgn-Poisson and 4-wgn-Matérn below). Its proof is hence skipped.

**Remark 1.**

- These theorems are linked to Poltyrev's capacity theorem [17].

- There exist stationary ergodic processes which are not capacity achieving for WGN, namely for which the property $\limsup_{k\to\infty} R_{n_k} < \frac{1}{2}\ln\frac{1}{2\pi e\sigma^2}$ is not sufficient for the probability of error to tend to 0 with $n$. For instance, consider a stationary cubic grid with points at

$$U + k(1)e^{-R}e(1) + \cdots + k(n)e^{-R}e(n),$$

where $e(1), \ldots, e(n)$ is the orthonormal basis of $\mathbb{R}^n$, $k(1), \ldots, k(n)$ range over the integers and $U$ is uniform over the $n$-cube centered in the origin and with side $e^{-R}$ (here $R \in \mathbb{R}$). The Voronoi cell of any point is the $n$-cube centered at this point and of side $e^{-R}$, with volume $e^{-nR}$. Hence the intensity of this point process is $\lambda_n = e^{nR}$. In the white Gaussian noise case,

$$p_s(n) = \left( \int_{-\frac{e^{-R}}{2}}^{\frac{e^{-R}}{2}} \frac{1}{\sqrt{2\pi\sigma^2}} \exp\left(-\frac{x^2}{2\sigma^2}\right) dx \right)^n.$$

So for all $R$ and $\sigma$, $p_s(n) \to 0$ when $n \to \infty$.

## 5.2 White Noise under Typicality Decoding

We assume now that each displacement vector has i.i.d. coordinates with a general distribution, the displacement vectors being i.i.d. from point to point. Let $D$ be a typical coordinate r.v. We assume that $D$ has a density $f$ with well defined differential entropy

$$h(D) = -\int_\mathbb{R} f(x) \ln(f(x)) dx.$$

Let

$$c(D) = -h(D). \tag{5}$$

### 5.2.1 Main Results

Our two main results are:

**Theorem 1-wn.** *For all point processes $\mu^n$ such that there is a subsequence $n_k \to \infty$ with $\liminf_{k \to \infty} R_{n_k} > c(D)$, for all choices of decoding regions $\mathcal{C}_k^n$ which are subsets of $\mathbb{R}^n$ jointly stationary with the points and the displacements and forming a decomposition of $\mathbb{R}^n$, we have $\lim_{k \to \infty} p_e(n_k) = 1$.* □

**Theorem 2-wn.** *Let $\mu^n$ be a Poisson point process of intensity $\lambda_n = e^{nR_n}$. For any subsequence $n_k \to \infty$ such that $\limsup_{k \to \infty} R_{n_k} < c(D)$, it is possible to choose decoding regions $\mathcal{C}_k^n$ that are subsets of $\mathbb{R}^n$ jointly stationary with the points and the displacements, forming a decomposition of $\mathbb{R}^n$, such that $\lim_{k \to \infty} p_e(n_k) = 0$.* □

**Remark 2.**

- Together these results give a kind of capacity theorem for stationary point process perturbed by additive noise: no point process with an error probability tending to zero offers a rate (normalized logarithmic density) larger than $c(D)$. This is related to the concept of *information-theoretic sphere packing* coined by Loeliger [13].

- Theorem 1-wn specialized to the case of i.i.d. Gaussian noise generalizes Theorem 1-wgn to arbitrary decoding functionals as defined in Section 3.2.

- The Poisson point process is capacity-achieving for white noise in that it allows for an error probability tending to zero for any rate smaller than $c(D)$.

We now prove these two theorems. The proofs rely on the following *typicality sets*: for all $n$ and $\delta > 0$, let

$$\mathcal{A}_\delta^n = \left\{ (x(1),\ldots,x(n)) \in \mathbb{R}^n \; : \; \left| -\frac{1}{n}\sum_{i=1}^n \log f(x(i)) - h(D) \right| < \delta \right\}. \quad (6)$$

By the SLLN,

$$\mathbb{P}_0^n((D_0^n(1),\ldots,D_0^n(n)) \in \mathcal{A}_\delta^n) \to_{n\to\infty} 1.$$

**Proof of Theorem 1-wn.** For all decompositions $\{\mathcal{C}_k^n\}_k$, we have

$$\mathbb{P}_0^n(D_0^n \in \mathcal{C}_0^n) \leq \mathbb{E}_0^n(1_{D_0^n \in \mathcal{C}_0^n \cap \mathcal{A}_\delta^n}) + \mathbb{E}_0^n(1_{D_0^n \notin \mathcal{A}_\delta^n}).$$

The second term tends to 0 as $n$ tends to infinity because of the SLLN. The first term is

$$\mathbb{E}_0^n \left( \int_{\mathcal{C}_0^n \cap \mathcal{A}_\delta^n} \prod_{i=1}^n f(x(i))dx(i) \right).$$

It is bounded from above by

$$e^{-n(h(D)-\delta)} \mathbb{E}_0^n \left( \mathrm{Vol}(\mathcal{C}_0^n) \right).$$

But for all translation invariant decompositions of the Euclidean space

$$\mathbb{E}_0^n \left( \mathrm{Vol}(\mathcal{C}_0^n) \right) = \frac{1}{\lambda_n} = e^{-nR_n}.$$

Hence the first term is bounded from above by

$$e^{-nR_n} e^{-n(h(D)-\delta)},$$

which tends to 0 when $n = n_k$ tends to infinity under the assumptions of the theorem.

**Proof of Theorem 2-wn.** Let $\{\mathcal{C}_k^n\}_k$ be the following decomposition of $\mathbb{R}^n$:

$$\begin{aligned}
\mathcal{C}_k^n &= \{(T_k^n + \mathcal{A}_\delta^n) \cap \{\cup_{l\neq k}(T_l^n + \mathcal{A}_\delta^n)\}^c\} \\
&\bigcup \{\mathcal{V}_k^n \cap \{\cup_{l\neq l'} [(T_l^n + \mathcal{A}_\delta^n) \cap (T_{l'}^n + \mathcal{A}_\delta^n)]\}\} \\
&\bigcup \{\mathcal{V}_k^n \cap \{\cup_l (T_l^n + \mathcal{A}_\delta^n)^c\}\},
\end{aligned}$$

where $\mathcal{V}_k^n$ denotes the Voronoi cell of $T_k^n$. In words, $\mathcal{C}_n^k$ contains

- all the locations $x$ which belong to the set $T_k^n + \mathcal{A}_\delta^n$ and to no other set of the form $T_l^n + \mathcal{A}_\delta^n$;

- all the locations $x$ that are ambiguous (i.e. belong to two or more such sets) and which are closer to $T_k^n$ than to any other point;
- all the locations which are uncovered (i.e. belong to no such set) and which are closer to $T_k^n$ than to any other point.

This scheme will be referred to as *typicality decoding* in what follows.

Let $\mu_!^n = \mu^n - \epsilon_0$ under $\mathbb{P}_0^n$. We use the following bound:

$$\mathbb{P}_0^n(D_0^n \notin \mathcal{C}_0^n) \leq \mathbb{P}_0^n(D_0^n \notin \mathcal{A}_\delta^n) + \mathbb{P}_0^n(D_0^n \in \mathcal{A}_\delta^n, \mu_!^n(D_0^n - \mathcal{A}_\delta^n) > 0).$$

The first term tends to 0 because of the SLLN. For the second, we use Slivnyak's theorem to bound it from above by

$$\mathbb{P}^n(\mu^n(D_0^n - \mathcal{A}_\delta^n) > 0) \leq \mathbb{E}^n(\mu^n(D_0^n - \mathcal{A}_\delta^n)) = \mathbb{E}^n(\mu^n(-\mathcal{A}_\delta^n)) = e^{nR_n}|\mathcal{A}_\delta^n|.$$

But

$$\begin{aligned} 1 &\geq \mathbb{P}^n(D_0^n \in \mathcal{A}_\delta^n) = \int_{\mathcal{A}_\delta^n} \prod_{i=1}^n f(x(i))dx(i) = \int_{\mathcal{A}_\delta^n} e^{n\frac{1}{n}\sum_{i=1}^n \ln(f(x(i)))}dx \\ &\geq \int_{\mathcal{A}_\delta^n} e^{n(-h(D)-\delta)}dx = e^{-n(h(D)+\delta)}|\mathcal{A}_\delta^n|, \end{aligned}$$

so that

$$|\mathcal{A}_\delta^n| \leq e^{n(h(D)+\delta)}.$$

Hence the second term is bounded from above by $e^{nR_n}e^{n(h(D)+\delta)}$. Using this for $n = n_k$ allows one to conclude the proof. □

### 5.2.2 Examples

Here are a few examples of geometrical shapes for the typicality sets.

**Gaussian.** In the Gaussian case, $\ln(f(x)) = -\frac{1}{2}\ln(2\pi\sigma^2) - \frac{x^2}{2\sigma^2}$ and $h(D) = \frac{1}{2}\ln(2\pi e\sigma^2)$, so that

$$\begin{aligned} \mathcal{A}_\delta^n &= \left\{ x \in \mathbb{R}^n : \left| \frac{1}{2}\ln(2\pi\sigma^2) + \frac{1}{2n\sigma^2}\sum_{i=1}^n x(i)^2 - \frac{1}{2}\ln(2e\pi\sigma^2) \right| < \delta \right\} \\ &= \left\{ x \in \mathbb{R}^n : \left| \sum_{i=1}^n x(i)^2 - n\sigma^2 \right| < 2n\delta\sigma^2 \right\}, \end{aligned}$$

with $x = (x(1), \ldots, x(n))$. That is $\mathcal{A}_\delta^n$ is the difference of two concentric balls of radius appr. $\sqrt{n}\sigma$.

**Remark 3.** Assuming WGN, the decoding regions of Theorem 2-wn and Theorem 2-wgn do not coincide. Theorem 2-wn does not admit Theorem 2-wgn as an immediate corollary in the Gaussian case since the decoding regions are not the same.

**Remark 4.** Let us stress that in Theorem 2-wn, the property $p_e(n) \to_{n \to \infty} 0$ does not require that
$$\mathbb{P}_0^n(\mathcal{A}_\delta^n \subset \mathcal{V}_0^n) \to 1$$
as $n \to \infty$. For example, in the WGN case, the last property does not true for all Poisson point processes with an intensity less than capacity. To see this, use the fact that $x$ belongs to $\mathcal{V}_0^n$ iff the open ball $B^n(x, |x|)$ contains no point of $\mu^n$. This implies that

$$\begin{aligned}
\mathbb{P}_0^n(\mathcal{A}_\delta^n \subset \mathcal{V}_0^n) &= \mathbb{P}_0^n\left(\cap_{x \in S^{n-1}(\sqrt{n}\sigma\sqrt{1+2\delta})} \{\mu^n(B^n(x, |x|)) = 0\}\right) \\
&= \mathbb{P}_0^n\left(\mu^n(B^n(0, 2\sqrt{n}\sigma\sqrt{1+2\delta})) = 1\right) \\
&= \mathbb{P}^n\left(\mu^n(B^n(0, 2\sqrt{n}\sigma\sqrt{1+2\delta})) = 0\right) \\
&= \exp(-\lambda_n V_B^n(2\sqrt{n}\sigma\sqrt{1+2\delta})),
\end{aligned}$$

where the penultimate identity follows from Slivnyak's theorem. It is easy to show (see Equation (131) in Appendix 10.3) that if $\lambda_n = e^{nR}$, where $R = \frac{1}{2}\ln\frac{1}{2\pi e\alpha^2\sigma^2}$ for $\alpha > 1$, then

$$\frac{C}{\sqrt{n+2}}\left(\frac{2\sqrt{1+2\delta}}{\alpha}\right)^n \leq \lambda_n V_B^n(2\sqrt{n}\sigma\sqrt{1+2\delta})) \leq \frac{C'}{\sqrt{n+2}}\left(\frac{2\sqrt{1+2\delta}}{\alpha}\right)^n,$$

for some constants $C$, and $C'$. Hence if $\alpha \geq 2\sqrt{1+2\delta}$,
$$\mathbb{P}_0^n(\mathcal{A}_\delta^n \subset \mathcal{V}_0^n) \to 1$$
whereas if $\alpha < 2\sqrt{1+2\delta}$,
$$\mathbb{P}_0^n(\mathcal{A}_\delta^n \subset \mathcal{V}_0^n) \to 0$$
as $n \to \infty$.

**Symmetric Exponential.** Consider the case where $D$ follows the symmetric exponential density
$$f(x) = \frac{\sqrt{2}}{2\sigma}\exp\left(-\frac{|x|\sqrt{2}}{\sigma}\right), \quad x \in \mathbb{R}.$$

This law is centered and with variance $\sigma^2$ and $E(|D|) = \sigma/\sqrt{2}$. The differential entropy of this density is $h(D) = \ln(\sqrt{2}e\sigma)$ and since $\ln(f(x)) = -\ln(\sqrt{2}\sigma) - \frac{|x|\sqrt{2}}{\sigma}$,

$$\begin{aligned}
\mathcal{A}_\delta^n &= \left\{(x(1),\ldots,x(n)) \in \mathbb{R}^n : \left|\ln(\sqrt{2}\sigma) + \frac{\sqrt{2}}{n\sigma}\sum_{i=1}^n |x(i)| - \ln(\sqrt{2}e\sigma)\right| < \delta\right\} \\
&= \left\{(x(1),\ldots,x(n)) \in \mathbb{R}^n : \left|\sum_{i=1}^n |x(i)| - \frac{n\sigma}{\sqrt{2}}\right| < n\delta\frac{\sigma}{\sqrt{2}}\right\}
\end{aligned}$$

i.e. $\mathcal{A}_\delta^n$ is the difference of two concentric $L_1$ balls of radius appr. $n\sigma/\sqrt{2}$.

**Uniform.** Let $D$ be uniform on $[-\sqrt{3}\sigma, +\sqrt{3}\sigma]$, which is again centered and with variance $\sigma^2$. Its differential entropy is $h(D) = \ln(2\sqrt{3}\sigma)$. Hence

$$\begin{aligned}\mathcal{A}^n_\delta &= \left\{x \in \mathbb{R}^n \,:\, \left|\ln(2\sqrt{3}\sigma) - \frac{1}{n}\sum_{i=1}^n \ln(1_{|x(i)| \leq \sqrt{3}\sigma}) - \ln(2\sqrt{3}\sigma)\right| < \delta\right\} \\ &= \left\{x \in \mathbb{R}^n \,:\, |x(i)| \leq \sqrt{3}\sigma) \text{ for all } i\right\}\end{aligned}$$

i.e. $\mathcal{A}^n_\delta$ is the $n$-cube of side $2\sqrt{3}\sigma$.

## 5.3 Stationary and Ergodic Noise under Typicality Decoding

A natural generalization of both § 5.1 and § 5.2 and is that where the displacement vectors are independent of $\mu^n$ and i.i.d., but the typical displacement vector $D^n$ has coordinates which form a stationary and ergodic (rather than i.i.d.) sequence of real-valued random variables. As already mentioned, we assume that for all $n$, $D^n = (\mathcal{D}_1, \ldots, \mathcal{D}_n)$, with $\{\mathcal{D}_k\}_k$ a real-valued, centered, stationary and ergodic stochastic process. We assume displacement vectors $D^n$ with a density $f^n$ admitting a differential entropy rate $h(\mathcal{D})$ and we define

$$c(\mathcal{D}) = -h(\mathcal{D}) \tag{7}$$

Here are two examples.

- **Colored Gaussian noise (cgn).** Assume that $\{\mathcal{D}_k\}$ is a regular stationary and ergodic Gaussian process with spectral density function $g(\beta)$, i.e.

$$\mathbb{E}(\mathcal{D}_0 \mathcal{D}_k) = \frac{1}{2\pi}\int_{-\pi}^{\pi} e^{ik\beta} g(\beta) d\beta,$$

for all $k$. Let $\Gamma_n$ be the $n \times n$ matrix with entries

$$\Gamma_n(i,j) = \mathbb{E}(\mathcal{D}_i \mathcal{D}_j) = r_{|j-i|}.$$

Let $f^n$ denote the Gaussian density on $\mathbb{R}^n$ with mean 0 and covariance matrix $\Gamma_n$, i.e.

$$f^n(x^n) = \frac{1}{(2\pi)^{n/2}\mathrm{Det}(\Gamma_n)^{1/2}} \exp\left(-\frac{1}{2}(x^n)^t \Gamma_n^{-1} x^n\right). \tag{8}$$

It is well known (see e.g. [10]) that the *differential entropy rate* of such a stationary process, namely

$$h(\mathcal{D}) = -\lim_{n \to \infty} \frac{1}{n}\int_{\mathbb{R}^n} f^n(x^n) \ln(f^n(x^n)) dx^n,$$

exists and is given by

$$\begin{aligned}h(\mathcal{D}) &= \frac{1}{2}\ln\left(2e\pi \exp\left(\frac{1}{2\pi}\int_{-\pi}^{\pi} \ln(g(\beta))d\beta\right)\right) \\ &= \frac{1}{4\pi}\int_{-\pi}^{\pi} \ln(2e\pi g(\beta))d\beta. \end{aligned} \tag{9}$$

- **Markov noise (mn).** Here we assume that $\{\mathcal{D}_n\}$ is a stationary Markov chain with values in $\mathbb{R}$, stationary distribution $\pi(x)dx$ with mean 0 and with transition kernel $P(dy \mid x) = p(y \mid x)dy$, where $p(y \mid x)$ is a density on $\mathbb{R}$. Then, when assuming that $(\mathcal{D}_1, \mathcal{D}_2)$ has a well defined differential entropy, we get that

$$\begin{aligned} h(\mathcal{D}) &= \lim_{n\to\infty} -\frac{1}{n} E\left(\ln(\pi(\mathcal{D}_1) P(\mathcal{D}_2|\mathcal{D}_1) \cdots P(\mathcal{D}_n|\mathcal{D}_{n-1}))\right) \\ &= -E\left(\ln(P(\mathcal{D}_2|\mathcal{D}_1))\right) \\ &= -\int_{\mathbb{R}^2} \pi(x) p(y \mid x) \ln(p(y \mid x)) dx dy = h(\mathcal{D}_2|\mathcal{D}_1) \;, \end{aligned}$$

    with $h(U|V)$ the conditional entropy of $V$ given $U$.

Let

$$\mathcal{A}_\delta^n = \left\{ x^n = (x(1), \ldots, x(n)) \in \mathbb{R}^n \;:\; \left| -\frac{1}{n} \log(f^n(x^n)) - h(\mathcal{D}) \right| < \delta \right\}. \quad (10)$$

**Theorem 1-sen.** *For all point processes $\mu^n$ such that there is a subsequence $n_k \to \infty$ with $\liminf_{k\to\infty} R_{n_k} > c(\mathcal{D})$, for all choices of decoding regions $\mathcal{C}_k^n$ which are subsets of $\mathbb{R}^n$ jointly stationary with the points and forming a decomposition of $\mathbb{R}^n$, we have $\lim_{k\to\infty} p_e(n_k) = 1$.* □

**Theorem 2-sen.** *Let $\mu^n$ be a Poisson point process of intensity $\lambda_n = e^{nR_n}$. For any subsequence $n_k \to \infty$ such that $\limsup_{k\to\infty} R_{n_k} < c(\mathcal{D})$, it is possible to choose decoding regions $\mathcal{C}_k^n$ that are subsets of $\mathbb{R}^n$ jointly stationary with the points and the displacements, forming a decomposition of $\mathbb{R}^n$, such that $\lim_{k\to\infty} p_e(n_k) = 0$.* □

**Proofs.** The proofs are verbatim extensions of Theorems 1-wn and 2-wn when using

- typicality sets (10) in place of (6);
- the ergodic assumption on the underlying displacement process and the pointwise ergodic theorem in place of the SLLN.

□

Notice that in the Gaussian case,

$$\mathcal{A}_\delta^n = \left| \frac{1}{n} (x^n)^t \Gamma_n^{-1} x^n - 1 + d(n) \right| < 2\delta, \quad (11)$$

with

$$d(n) = \frac{1}{n} \ln(\text{Det}(\Gamma_n)) - \left( \frac{1}{2\pi} \int_{-\pi}^{\pi} \ln(g(\beta)) d\beta \right) \to_{n\to\infty} 0.$$

The last limit follows from the Grenander–Szegö Theorem ([10]).

# 6 Palm Probability Representations of the Error Probability under MLE

The setting is that of Section 3. The following representations of the error probability will be used for the evaluation of the error exponents in the next section.

## 6.1 White Gaussian Noise

**Theorem 3-wgn.** *For all stationary and ergodic point processes and all i.i.d. white Gaussian displacement vectors, the probability of success under MLE (or equivalently when $\mathcal{C}_0^n$ is the Voronoi cell of 0 w.r.t. $\mu^n$) is*

$$p_s(n) = \int_{r \geq 0} \int_{\vec{v} \in SS^{n-1}} \mathbb{P}_0^n(\mu^n(B^n(r\vec{v}, r)) = 0) \frac{g_\sigma^n(r)}{A_{n-1}} d\vec{v} dr , \qquad (12)$$

*with $\mathbb{P}_0^n$ the Palm probability of $\mu^n$ and with*

$$g_\sigma^n(r) = 1_{r > 0} e^{-\frac{r^2}{2\sigma^2}} \frac{1}{2^{n/2}} \frac{r^{n-1}}{\sigma^n} \frac{2}{\Gamma(n/2)}. \qquad (13)$$

*If $\mu^n$ is isotropic (invariant by rotations), then*

$$p_s(n) = \int_{r \geq 0} \mathbb{P}_0^n(\mu^n(B^n(r\vec{v}, r)) = 0) g_\sigma^n(r) dr , \qquad (14)$$

*where $\vec{v}$ is any unit vector of $\mathbb{R}^n$. If $\mu^n$ is a Poisson point process, then*

$$p_s(n) = \int_0^\infty e^{-\lambda_n V_B^n(r)} g_\sigma^n(r) dr = \int_0^\infty e^{-\lambda_n V_B^n(r\sigma)} g_1^n(r) dr , \qquad (15)$$

*with $V_B^n(r)$ the volume of the ball $B^n(0, r)$.* □

**Proof.** Using the fact that $x$ belongs to the Voronoi cell $\mathcal{V}_0^n$ of point 0 iff the open ball $B^n(x, |x|)$ contains no point of $\mu^n$, we get from (2) that

$$p_s(n) = \mathbb{P}_0^n(D_0^n \in \mathcal{V}_0^n) = \mathbb{P}_0^n(\mu^n(B^n(D_0^n, |D_0^n|)) = 0). \qquad (16)$$

Since $D_0^n$ is a Gaussian vector with i.i.d. components, its norm $|D_0^n|$ has for density $g_\sigma^n(x) = g_1^n(x/\sigma)/\sigma$ on $\mathbb{R}^+$ (see Appendix 10.1). In addition, given that $|D_0^n| = r$, the law of the angle is the uniform law on $SS_{n-1}$.

In the Poisson case, from Slivnyak's theorem,

$$\mathbb{P}_0^n(\mu^n(B^n(r\vec{v}, r)) = 0) = \mathbb{P}^n(\mu^n(B^n(r\vec{v}, r)) = 0) = e^{-\lambda_n V_B^n(r)}.$$

□

## 6.2 Colored Gaussian Noise

In the colored Gaussian noise case, consider the orthonormal basis which diagonalizes the covariance matrix $\Gamma_n$. In this basis,

$$(x^n)^t \Gamma_n^{-1} x^n = \sum_{i=1}^n \sigma_i^{-2} x_i^2,$$

i.e. the coordinates of the noise are independent and coordinate $i$ is Gaussian, centered and with variance $\sigma_i^2$ with $\sigma_i$ the $i$-th eigenvalue of $\Gamma_n$.

Let $\mathcal{R}(\mu^n)$ be the image of the point process $\mu^n$ by this change of basis. Clearly, $\mathcal{R}(\mu^n)$ is also stationary, but its law may be different from that of $\mu^n$.

For $x^n$ and $y^n$ two vectors of $\mathbb{R}^n$, let

$$d^2(x^n, y^n) = \sum_{i=1}^n \sigma_i^{-2}(y_i - x_i)^2.$$

Let $\mathcal{E}^n(x^n, r)$ denote the ellipsoid:

$$\mathcal{E}^n(x^n, r) = \{y^n \in \mathbb{R}^n \text{ s.t. } d(x^n, y^n) < r\}. \tag{17}$$

Note that the volume of this ellipsoid only depends on $r$ and we will denote it by $V_\mathcal{E}^n(r)$. We have:

**Theorem 3-cgn.** *For all stationary and ergodic point processes $\mu^n$ and all colored Gaussian displacement vectors, the probability of success under MLE (i.e. that with $\mathcal{C}_0^n$ equal to the Voronoi cell of point 0 w.r.t. $\mu^n$ and for the distance $d$) is*

$$p_s(n) = \int_{x^n \in \mathbb{R}^n} \mathbb{Q}_n^0(\mathcal{R}(\mu^n)(\mathcal{E}^n(x^n, d(x^n, 0))) = 0) \prod_{i=1}^n \frac{e^{-\frac{x_i^2}{2\sigma_i^2}}}{\sqrt{2\pi\sigma_i^2}} dx^n, \tag{18}$$

*where $\mathbb{Q}_n^0$ denotes the Palm probability w.r.t. $\mathcal{R}(\mu^n)$. If $\mu^n$ is isotropic, then $\mu^n$ and $\mathcal{R}(\mu^n)$ have the same law so that*

$$p_s(n) = \int_{x^n \in \mathbb{R}^n} \mathbb{P}_0^n(\mu^n(\mathcal{E}^n(x^n, d(x^n, 0))) = 0) \prod_{i=1}^n \frac{e^{-\frac{x_i^2}{2\sigma_i^2}}}{\sqrt{2\pi\sigma_i^2}} dx^n. \tag{19}$$

*If $\mu^n$ is Poisson, then*

$$p_s(n) = \int_{r>0} \exp\left(-\lambda_n V_\mathcal{E}^n(r)\right) g_1^n(r) dr, \tag{20}$$

*where the last function is defined in (13).* □

**Proof.** The first property follows from the independence of the noise and the point process and from the MLE decoding rule which states that when $x^n$ is received, one returns the codeword

$$\operatorname{argmin}_k \, d(x^n, T_k^n).$$

So the probability of success (given that 0 is sent and that the additive noise is $x^n$) is the probability that $\mu^n$ has no point $T_k^n \neq 0$ such that $d(x^n, T_k^n) < d(x^n, 0)$. But

$$d(x^n, T_k^n) > d(x^n, 0), \, \forall k \neq 0 \quad \Leftrightarrow \quad \mu^n(\mathcal{E}^n(x^n, d(x^n, 0))) = 0$$

and this concludes the proof of the first statement.

If the point process is isotropic, then $\mathbb{Q}_0^n = \mathbb{P}_0^n$.

For the Poisson case, we use (19) and then Slivnyak's theorem and stationarity, which give that for all $x^n$ such that $d(x^n, 0) = r$,

$$\mathbb{P}_0^n(\mu^n(\mathcal{E}^n(x^n, r)) = 0) = \mathbb{P}^n(\mu^n(\mathcal{E}^n(x^n, r)) = 0) = \mathbb{P}^n(\mu^n(\mathcal{E}^n(0, r)) = 0).$$

The proof is then concluded when noting that the density of $d(D^n, 0)$ is $g_1^n$. □

## 6.3 Stationary and Ergodic Noise under MLE

We consider here the case alluded to above where the noise $D^n$ has its coordinates which are the $n$ first values of a discrete time, stationary and ergodic, real–valued stochastic process $\{\mathcal{D}_k\}_{k \in \mathbb{Z}}$. We assume that the latter is centered and we assume that $D^n$ has a density for all $n$ denoted by $f^n$. We will call *stun* of $f^n$ the function $y^n \in \mathbb{R}^n \to -\ln(f^n(y^n)) \in \mathbb{R}$, which measures one's astonishment of getting the sample $y^n$ when sampling from $f^n$. We define the *stun discrepancy* of $s^n \in \mathbb{R}^n$ w.r.t. $t^n \in \mathbb{R}^n$ as

$$\mathbb{D}(s^n, t^n) = -\frac{1}{n} \ln(f^n(s^n - t^n)). \tag{21}$$

In the Gaussian case, $\mathbb{D} = K + d^2/n$, where $d$ is the distance defined above and $K$ is a constant. Note that $\mathbb{D}$ is not a distance in general. In particular it is neither symmetrical nor positive in general.

By definition, under MLE decoding, when $x^n$ is received, one returns the codeword

$$\operatorname{argmin}_k \mathbb{D}(x^n, T_k^n) \,,$$

provided the latter is uniquely defined (we will see below what happens if it is not the case).

Given that $0 = T_0^n$ is sent and that the additive noise is $x^n$, a sufficient condition for MLE decoding to be successful is that $\mu^n$ has no point $T_k^n \neq 0$ such that $f^n(x^n - T_k^n) \geq f^n(x^n)$ or equivalently such that $\mathbb{D}(x^n, T_k^n) \leq \mathbb{D}(x^n, 0)$. But for all $x^n$,

$$\mathbb{D}(x^n, T_k^n) > \mathbb{D}(x^n, 0), \, \forall k \neq 0 \quad \Leftrightarrow \quad (\mu^n - \epsilon_0)(\mathbb{F}(x^n)) = 0, \tag{22}$$

with
$$\begin{aligned}
\mathbb{F}(x^n) &= \{y^n \in \mathbb{R}^n \text{ s.t. } \mathbb{D}(x^n, y^n) \leq \mathbb{D}(x^n, 0)\} \\
&= \{y^n \in \mathbb{R}^n \text{ s.t. } -\frac{1}{n}\ln(f^n(x^n - y^n)) \leq -\frac{1}{n}\ln(f^n(x^n))\} \,. \quad (23)
\end{aligned}$$

Hence

$$p_s(n) \geq \mathbb{P}_0^n\left(\bigcap_{k \neq 0}\{\mathbb{D}(D^n, T_k^n) > \mathbb{D}(D^n, 0)\}\right) = \mathbb{P}_0^n\left((\mu^n - \epsilon_0)(\mathbb{F}(D^n)) = 0\right) \quad (24)$$

and

$$p_e(n) \leq \mathbb{P}_0^n\left(\cup_{k \neq 0}\{\mathbb{D}(D^n, T_k^n) \leq \mathbb{D}(D^n, 0)\}\right) = \mathbb{P}_0^n\left((\mu^n - \epsilon_0)(\mathbb{F}(D^n)) > 0\right). \quad (25)$$

Also notice that the volume of the set $\mathbb{F}(x^n)$ only depends on $\mathbb{D}(x^n, 0) = -\frac{1}{n}\ln(f^n(x^n))$: if this last quantity is equal to $u$, the associated volume is

$$\begin{aligned}
&\text{Vol}\left\{y^n \in \mathbb{R}^n \text{ s.t. } -\frac{1}{n}\ln(f^n(x^n - y^n)) \leq u\right\} \\
&= \text{Vol}\left\{y^n \in \mathbb{R}^n \text{ s.t. } -\frac{1}{n}\ln(f^n(-y^n)) \leq u\right\} \\
&= \text{Vol}\left\{y^n \in \mathbb{R}^n \text{ s.t. } -\frac{1}{n}\ln(f^n(y^n)) \leq u\right\}.
\end{aligned}$$

Hence this volume is then

$$W_{\mathcal{D}}^n(u) = \text{Vol}\left\{y^n \in \mathbb{R}^n \text{ s.t. } -\frac{1}{n}\ln(f^n(y^n)) \leq u\right\}, \quad u \in \mathbb{R} \cup \{\infty\}. \quad (26)$$

Notice that we have

$$\text{Vol}(\mathbb{F}(x^n)) = W_{\mathcal{D}}^n\left(-\frac{1}{n}\ln(f^n(x^n))\right). \quad (27)$$

**Theorem 3-sen.** *For all stationary and ergodic point processes $\mu^n$ and all i.i.d. displacement vectors, the probability of success under MLE is*

$$p_s(n) \geq \int_{x^n \in \mathbb{R}^n} \mathbb{P}_0^n((\mu^n - \epsilon_0)(\mathbb{F}(x^n)) = 0) f^n(x^n) dx^n \,. \quad (28)$$

*If $\mu^n$ is such that, under $\mathbb{P}_0^n$, the point process $\mu^n - \epsilon_0$ admits an intensity bounded from above by the function $g^n(.)$ on $\mathbb{R}^d$, then*

$$p_e(n) \leq \int_{x^n \in \mathbb{R}^n} \min\left(1, \int_{y^n \in \mathbb{R}^n} 1_{\mathbb{D}(x^n, y^n) \leq \mathbb{D}(x^n, 0)} g^n(y^n) dy^n\right) f^n(x^n) dx^n \,. \quad (29)$$

*If $\mu^n$ is Poisson of intensity $\lambda_n$, then*

$$p_s(n) \geq \int_{u \in \mathbb{R}} \exp\left(-\lambda_n W_{\mathcal{D}}^n(u)\right) \rho_{\mathcal{D}}^n(du) \,, \quad (30)$$

*where $\rho_{\mathcal{D}}^n(du)$ is the entropy spectrum of $f^n$, namely the law of the random variable $-\frac{1}{n}\ln(f^n(D^n))$ on $\mathbb{R}$.* □

**Proof.** The arguments for proving the first and last relations are as in the previous particular instances of noise processes. In order to prove (29), we use (25) to write:

$$\begin{aligned}
p_e(n) &\leq \mathbb{P}_0^n((\mu^n - \epsilon_0)(\mathbb{F}(D_0^n)) > 0) \\
&\leq \min\left(1, \mathbb{E}_0^n[(\mu^n - \epsilon_0)(\mathbb{F}(D_0^n))]\right) \\
&= \int_{x^n \in \mathbb{R}^n} f^n(x^n) \min\left(1, \mathbb{E}_0^n\left[\sum_{k \neq 0} 1_{\mathbb{D}(x^n, T_k^n) \leq \mathbb{D}(x^n, 0)}\right]\right) \\
&\leq \int_{x^n \in \mathbb{R}^n} f^n(x^n) \min\left(1, \int_{y^n \in \mathbb{R}^n} 1_{\mathbb{D}(x^n, y^n) \leq \mathbb{D}(x^n, 0)}\, g^n(y^n) dy^n\right) dx^n,
\end{aligned}$$

where we used the assumption on the intensity of $\mu^n - \epsilon_0$ under $\mathbb{P}_0^n$ to derive the last line. $\square$

**Terminology, Key Definitions and Properties.**

1. The real-valued random variable $-\frac{1}{n}\ln(f^n(D^n))$ is referred to as the *normalized entropy density* of $D^n$ [11].

2. The law of this real-valued random variable, $\rho_\mathcal{D}^n(du)$, is referred to as the *entropy spectrum* of $D^n$ [11]. Notice that:

   - The existence of a density for $\mathcal{D}^n$ does not imply that $\rho_\mathcal{D}^n(.)$ admits a density. The latter holds iff $\mathbb{P}^n(-\frac{1}{n}\ln(f^n(D^n)) \in B) = 0$ for all Borel sets $B$ with 0 Lebesgue measure. For instance when $f^n$ is constant and positive on a set of positive Lebesgue measure, the last condition does not hold.

   - The support of $\rho_\mathcal{D}^n(.)$ is not necessarily the whole real line. For instance, in the WGN case, the support is $[1/2\ln(2\pi\sigma^2), \infty)$. In the case white uniform noise (see Section 5.2.2), the support is the singleton $\{\ln(2\sqrt{3}\sigma)\}$.

3. The sets

$$\mathcal{S}_\mathcal{D}^n(u) = \{y^n \in \mathbb{R}^n \text{ s.t. } -\frac{1}{n}\ln(f^n(-y^n)) \leq u\}, \quad u \in \mathbb{R}, \qquad (31)$$

   will be referred to as the *stun level sets* of $D^n$.

4. The volume $W_\mathcal{D}^n(u)$ of $\mathcal{S}_\mathcal{D}^n(u)$ will be referred to as the *stun level volume* for $u$.

5. The measure $w_\mathcal{D}^n$ on $\mathbb{R}$ defined by:

$$w_\mathcal{D}^n(B) = \text{Vol}\{y^n \in \mathbb{R}^n \text{ s.t. } -\frac{1}{n}\ln(f^n(y^n)) \in B\}, \qquad (32)$$

   for all Borel sets $B$ of the real line, will be called the *stun level measure*. The following properties hold:

- The measures $w_\mathcal{D}^n$ and $\rho_\mathcal{D}^n$ are equivalent. Indeed, we have

$$\rho_\mathcal{D}^n(B) = \int_{-\frac{1}{n}\ln(f^n(x^n))\in B} f^n(x^n) dx^n \,,$$

which implies that for all bounded Borel sets $B$,

$$e^{-n\sup(B)} w_\mathcal{D}^n(B) \leq \rho_\mathcal{D}^n(B) \leq e^{-n\inf(B)} w_\mathcal{D}^n(B) \quad (33)$$
$$e^{n\inf(B)} \rho_\mathcal{D}^n(B) \leq w_\mathcal{D}^n(B) \leq e^{n\sup(B)} \rho_\mathcal{D}^n(B) \,. \quad (34)$$

Each measure is hence absolutely continuous w.r.t. the other (the absolute continuity condition is immediate for bounded sets in view of the last bounds and is then easily extended to arbitrary sets). The Radon–Nikodym derivative $\frac{dw_\mathcal{D}^n}{d\rho_\mathcal{D}^n}$ at $s \in \mathbb{R}$ is $e^{ns}$ and $\frac{d\rho_\mathcal{D}^n}{dw_\mathcal{D}^n}$ at $s$ is $e^{-ns}$.

- The measure $w_\mathcal{D}^n$ is $\sigma$-finite. This immediately follows from the last bound.

- For all $u$

$$W_\mathcal{D}^n(u) = \int_{(-\infty,u]} w_\mathcal{D}^n(ds) = \int_{(-\infty,u]} e^{ns} \rho_\mathcal{D}^n(ds) \,. \quad (35)$$

This is a direct corollary of the value of the above Radon–Nikodym derivative.

6. The *stun cell* $\mathcal{L}_k^n(\mathcal{D})$ of point $T_k^n$ is defined as follows:

$$\mathcal{L}_k^n(\mathcal{D}) = \{x^n \text{ s.t. } \mathbb{D}(x^n, T_k^n) < \inf_{l \neq k} \mathbb{D}(x^n, T_l^n)\} \quad (36)$$
$$\cup \ \{x^n \text{ s.t. } \mathbb{D}(x^n, T_k^n) = \mathbb{D}(x^n, T_l^n) \text{ for some } l \neq k\} \cap \mathcal{V}_k^n \,.$$

It is comprised of

- The locations $x^n$ with a stun discrepancy (w.r.t. $f^n$) to $T_k^n$ smaller than that to any other point;

- The locations $x^n$ with an ambiguous stun discrepancy (this includes the case where $\mathbb{D}(x^n, T_k^n) = \infty$ for all $k$) but which are closer to $T_k^n$ for Euclidean distance than to all other points of $\mu^n$.

These cells form a decomposition of the Euclidean space which will be referred to as the *stun cell decomposition* w.r.t. the point process $\mu^n$ for the noise $\mathcal{D}$ (more precisely $D^n$ or $f^n$). The stun cell decomposition w.r.t. any WGN noise with positive variance is the Voronoi decomposition: for all dimensions $n$, for all point processes $\mu^n$ on $\mathbb{R}^n$ and for all $k$, $\mathcal{L}_k^n(\text{WGN}) = \mathcal{V}_k^n$.

This definition is somewhat Gaussian-centric. Any other tessellation whose cells satisfy the conditions of §3 could be used in place of the Voronoi tessellation.

Here are a few observations on the presence of ambiguity. When there is no ambiguity, there is no need to introduce an exogenous tessellation. In addition, if
$$\mathrm{argmin}_k\{\mathbb{D}(D_0^n, T_k^n)\}$$
is $\mathbb{P}_0^n$-a.s. a singleton, (24) becomes an equality as well as (28) and (30). A (rough) sufficient condition for
$$\mathrm{argmin}_k\{\mathbb{D}(D_0^n, T_k^n)\}$$
to be a set of cardinality less than or equal to 1 is that for all $a^n \neq b^n \in \mathbb{R}^n$, the set
$$\{x^n \in \mathbb{R}^n \text{ s.t. } f^n(x^n - a^n) = f^n(x^n - b^n)\} \cap \mathrm{Support}(f^n) \qquad (37)$$
is of Lebesgue measure 0 in $\mathbb{R}^n$. Indeed, if this holds true, then for all $k \neq l$,
$$\mathbb{E}_0^n \int_{\mathbb{R}^n} 1_{f^n(x^n - T_k^n) = f^n(x^n - T_l^n)} f^n(x^n) dx^n = 0,$$
so that
$$\sum_{k \neq l} \mathbb{E}_0^n \int_{\mathbb{R}^n} 1_{f^n(x^n - T_k^n) = f^n(x^n - T_l^n)} f^n(x^n) dx^n = 0,$$
which in turn implies that
$$\mathbb{P}_0^n \left( \cup_{k \neq l} \{f^n(D^n - T_k^n) = f^n(D^n - T_l^n)\} \right) = 0.$$

**Remark 5.** Theorem 3-sen is consistent with the previous ones. For instance, in the wgn case, for all $x^n \in \mathbb{R}^n$,
$$-\frac{1}{n}\ln(f^n(x^n)) = \frac{1}{2}\ln(2\pi\sigma^2) + \frac{1}{2n\sigma^2}\sum_{i=1}^n x_i^2 . \qquad (38)$$

Using this, it is easy to show that (30) boils down to (15). This follows from the fact that for all $u \geq \frac{1}{2}\ln(2\pi\sigma^2)$:

- the set $\mathcal{S}_{\mathcal{D}}^n(u)$ is a $n$ ball of radius $\sqrt{2n\sigma^2\left(u - \frac{1}{2}\ln(2\pi\sigma^2)\right)}$, so that
$$W_{\mathcal{D}}^n(u) = V_B^n\left(\sqrt{2n\sigma^2\left(u - \frac{1}{2}\ln(2\pi\sigma^2)\right)}\right);$$

- the random variable $-\frac{1}{n}\ln(f^n(D^n))$ is equal to $\frac{1}{2}\ln(2\pi\sigma^2) + \frac{1}{2n\sigma^2}\sum_{i=1}^n D_i^2$, which implies that $\rho_{\mathcal{D}}^n(du)$ admits the density
$$g_\sigma^n\left(\sqrt{2n\sigma^2\left(u - \frac{1}{2}\ln(2\pi\sigma^2)\right)}\right) \frac{n\sigma^2}{\sqrt{2n\sigma^2\left(u - \frac{1}{2}\ln(2\pi\sigma^2)\right)}} 1_{u > \frac{1}{2}\ln(2\pi\sigma^2)}.$$

In order to derive (15) from (30), it is then enough to perform the change of variable $r = \sqrt{2n\sigma^2\left(u - \frac{1}{2}\ln(2\pi\sigma^2)\right)}$ in the latter.

### 6.4 Mismatched Decoding

In mismatched decoding, the decoder has been designed for some noise $\mathcal{D}$ but the actual noise is in fact $\widetilde{\mathcal{D}}$. By the same arguments as in Theorem 3-sen, we get:

**Theorem 3-sen-mismatch.** *For all stationary and ergodic point processes $\mu^n$ and all i.i.d. design displacement vectors $\mathcal{D}$ and actual displacement vectors $\widetilde{\mathcal{D}}$, the probability of error under MLE satisfies*

$$p_e(n) \leq \int_{x^n \in \mathbb{R}^n} \mathbb{P}_0^n((\mu^n - \epsilon_0)(\mathbb{F}(x^n)) > 0)\widetilde{f}^n(x^n)dx^n \ . \tag{39}$$

*If $\mu^n$ is Poisson of intensity $\lambda_n$, then*

$$p_e(n) \leq \int_{u \in \mathbb{R}} \left(1 - \exp\left(-\lambda_n W_{\mathcal{D}}^n(u)\right)\right) \rho_{\widetilde{\mathcal{D}}}^n(du) \ , \tag{40}$$

*where $\rho_{\widetilde{\mathcal{D}}}^n(du)$ is the law of the random variable $-\frac{1}{n}\ln(\widetilde{f}^n(\widetilde{D}^n))$ on $\mathbb{R}$ and $W_{\mathcal{D}}^n$ is the stun level volume for $\mathcal{D}$.* □

### 6.5 Matérn Point Process under MLE

A Matérn I point process is created by dropping points from a Poisson process as follows. Choose some positive radius called the exclusion radius. Any point in the initial Poisson process that has another point within this fixed radius of it is dropped (note that both points will be dropped since the first point will also be within the same fixed radius of the second point). This is the simplest type of *hard sphere exclusion*. For an information theorist, this is reminiscent of expurgation [8] and we will also use this term below to describe the transformation of the Poisson into the Matérn point process. A variant is the Matérn II processes, created from a Poisson processes as follows. Mark each point of the initial Poisson process with an independent real number chosen uniformly over the unit interval. Choose some positive radius. Every point of the original Poisson process examines the ball around it of this radius and survives only if its associated mark is strictly larger than the marks of all the other points in this ball. These processes were introduced by Matérn [14] (the Matérn II process will not be analyzed in this paper).

Note that neither Matérn I nor Matérn II are Poisson (this is because the thinning is not independent). However, both are isotropic.

Below, we introduce a new class of Matérn point processes meant to cope with the stationary and ergodic noise framework of Section 6.3. Assume for simplicity that $f^n(x^n) = f^n(-x^n)$. If two points $S$ and $T$ of the Poisson point process $\mu^n$ are such that $\mathbb{D}(T, S) < \xi$, with $\xi \in \mathbb{R}$ some threshold, then $T$ is discarded. The surviving points form the Matérn-$\mathcal{D}$-$\xi$ point process $\widehat{\mu}^n$.

**Theorem 3-sen-Matérn.** *Under the assumptions of Theorem 3-sen, the probability of error for the Matérn-$\mathcal{D}$-$\xi$ point process satisfies the bound*

$$p_e(n) \leq \int_{x^n \in \mathbb{R}^n} \min\left(1, \lambda_n \int_{y^n \in \mathbb{R}^n} 1_{\mathbb{D}(y^n,0) \geq \xi} \, 1_{\mathbb{D}(x^n,y^n) \leq \mathbb{D}(x^n,0)} \, dy^n\right) f^n(x^n) dx^n. \quad (41)$$

$\square$

**Proof** Let $\widehat{\mathbb{P}}_0^n$ denote the Palm probability of $\widehat{\mu}^n$. Under $\widehat{\mathbb{P}}_0^n$, the point process $\widehat{\mu}^n - \epsilon_0$ has an intensity bounded from above by $\lambda_n 1_{\mathbb{D}(y^n,0) \geq \xi}$ at $y^n$. The result then follows from (29). $\square$

Notice that the Matérn-WGN-$\xi$ model boils down to Matérn I for the exclusion radius

$$r_n(\xi) = \sqrt{2n\sigma^2}\sqrt{\xi - \frac{1}{2}\ln(2\pi\sigma^2)}, \quad (42)$$

for $\xi > \frac{1}{2}\ln(2\pi\sigma^2)$. The set $\{y^n : \mathbb{D}(y^n, 0) \geq \xi\}$ is then the complement of the ball $B^n(0, r_n(\xi))$ and for all $x^n \in \mathbb{R}^n$ with norm $r$, the set $\{y^n : \mathbb{D}(x^n, y^n) < \mathbb{D}(x^n, 0)\}$ is the ball $B^n(x^n, r)$. Hence

**Corollary 3-wgn-Matérn.** *In the WGN case,*

$$p_e(n) \leq \int_{r > 0} \min\left(1, \lambda_n \, Vol(B^n(0, r_n(\xi))^c \cap B^n(x^n(r), r))\right) g_\sigma^n(r) dr \quad (43)$$

*with $x^n(r) = (r, 0, \ldots, 0) \in \mathbb{R}^n$ and $r_n(.)$ defined in (42).* $\square$

**Proof** The result immediately follows from (38), (41) and (42). $\square$

In the general (sen) case, the unfortunate fact that the volume of the vulnerability set (the set which ought to be empty of points for no error to occur):

$$\int_{y^n \in \mathbb{R}^n} 1_{\mathbb{D}(y^n,0) \geq \xi} \, 1_{\mathbb{D}(x^n,y^n) \leq \mathbb{D}(x^n,0)} dy^n$$

now depends on the point $x^n$, and not only on the value of $\mathbb{D}(x^n, 0)$ as in the non-expurgated case, can be taken care of by introducing the following upper bound

$$M_\mathcal{D}^n(u, \xi) = \sup_{x^n: \, \mathbb{D}(x^n,0)=u} \int_{y^n \in \mathbb{R}^n} 1_{\mathbb{D}(y^n,0) \geq \xi} \, 1_{\mathbb{D}(x^n,y^n) \leq u} \, dy^n, \quad (44)$$

which only depends on $\mathbb{D}(x^n, 0)$. This quantity will be referred to as the *expurgated stun level volume*. By the same arguments as above, we get:

**Corollary 3-sen-Matérn.** *The assumptions on $\mathcal{D}$ are those of Theorem 3-sen. The probability of error for the Matérn-$\mathcal{D}$-$\xi$ point process satisfies the bound*

$$p_e(n) \leq \int_{u \in \mathbb{R}} \min\left(1, \lambda_n M_\mathcal{D}^n(u, \xi)\right) \rho_\mathcal{D}^n(du) \ . \tag{45}$$

□

Here are further bounds on $M_\mathcal{D}^n(u, \xi)$:

$$M_\mathcal{D}^n(u, \xi) \leq \int_{y^n \in \mathbb{R}^n} 1_{\mathbb{D}(y^n, 0) \geq \xi} \, 1_{\exists x^n: \, \mathbb{D}(x^n, 0) = u, \, \mathbb{D}(y^n, x^n) \leq u} \, dy^n \tag{46}$$

where we note the standing assumption in this section that $f^n(x^n) = f^n(-x^n)$, and

$$M_\mathcal{D}^n(u, \xi) \leq \int_{z^n \in \mathbb{R}^n} 1_{\mathbb{D}(z^n, 0) \leq u} \, 1_{\exists x^n: \, \mathbb{D}(x^n, 0) = u, \, \mathbb{D}(z^n + x^n, 0) \geq \xi} \, dz^n \ . \tag{47}$$

The first bound immediately follows from (44) and the fact that $\sup \int \leq \int \sup$. The second bound is obtained in the same way after the change of variable $z^n = y^n - x^n$ in (44). This second bound can be seen as a natural extension of (26) since

$$W_\mathcal{D}^n(u) = \int_{z^n \in \mathbb{R}^n: \, \mathbb{D}(z^n, 0) \leq u} dz^n$$

$$\geq \int_{z^n \in \mathbb{R}^n: \, \mathbb{D}(z^n, 0) \leq u} 1_{\exists x^n: \, \mathbb{D}(x^n, 0) = u, \, \mathbb{D}(z^n + x^n, 0) \geq \xi} \, dz^n \geq M_\mathcal{D}^n(u, \xi) \ .$$

# 7 Further Representations of the Error Probability

In this section, we come back to the general framework of Section 3 and we give some other Palm probability representations of the error probability:

$$p_e(n) = \mathbb{P}_0^n\left(D_0^n \notin \mathcal{C}_0^n\right),$$

## 7.1 Mass Transport Principle Representation

The results of this section hold for arbitrary decoding regions.
**Theorem 3.2-sen.** *Under the assumptions of Section 3,*

$$p_e(n) = \mathbb{E}_0^n\left(\left(\sum_{k \neq 0} \epsilon_{T_k^n + D_k^n}\right)(\mathcal{C}_0^n)\right) \ . \tag{48}$$

□

**Proof.** From the very definition of the Palm probability,

$$\begin{aligned}
\lambda_n \mathbb{P}_0^n \left( D_0^n \notin \mathcal{C}_0^n \right) &= \mathbb{E}^n \left( \sum_{i, T_i^n \in [0,1]^n} 1_{T_i^n + D_i^n \notin \mathcal{C}_i^n} \right) \\
&= \mathbb{E}^n \left( \sum_{i, T_i^n \in [0,1]^n} \sum_{v \in \mathbb{Z}^n} \sum_{j, T_j^n \in v + [0,1]^n} \kappa(i,j) \right) \\
&= \sum_{v \in \mathbb{Z}^n} \mathbb{E}^n \left( \sum_{i, T_i^n \in [0,1]^n} \sum_{j, T_j^n \in v + [0,1]^n} \kappa(i,j) \right),
\end{aligned}$$

with

$$\kappa(i,j) = 1_{T_i^n + D_i^n \in \mathcal{C}_j^n} 1_{T_i^n \neq T_j^n}.$$

But for all $v \in \mathbb{Z}^n$,

$$\mathbb{E}^n \left( \sum_{i, T_i^n \in [0,1]^n} \sum_{j, T_j^n \in v + [0,1]^n} \kappa(i,j) \right) = \mathbb{E}^n \left( \sum_{i, T_i^n \in -v + [0,1]^n} \sum_{j, T_j^n \in [0,1]^n} \kappa(i,j) \right)$$

Hence

$$\begin{aligned}
\lambda_n \mathbb{P}_0^n \left( D_0^n \notin \mathcal{C}_0^n \right) &= \mathbb{E}^n \left( \sum_{v \in \mathbb{Z}^n} \sum_{i, T_i^n \in -v + [0,1]^n} \sum_{j, T_j^n \in [0,1]^n} \kappa(i,j) \right) \\
&= \lambda_n \mathbb{E}_0^n \left( \left( \sum_{i \neq 0} \epsilon_{T_i^n + D_i^n} \right) (\mathcal{C}_0^n) \right),
\end{aligned}$$

where the last relation comes from the definition of Palm probability. □

Notice that the last proof only requires that the sequence $\{D_k^n, \mathcal{C}_k^n\}_k$ be a sequence of marks of the point process $\mu^n = \sum_k \epsilon_{T_k^n}$. The independence between the point process and the displacement is hence not required in Theorem 3.2-sen.

The next result requires the independence between the marked point process $\{T_k^n, \mathcal{C}_k^n\}_k$ and the displacements $\{D_k^n\}_k$. It uses the decoding functional framework of Section 3.2. For all point measures $\nu$ on $\mathbb{R}^n$ and all $x$ such that $x$ belongs to the support of $\nu$, let $\mathcal{C}_x^n(\nu) = \mathbb{C}^n(x, \nu)$ be the decoding region of $x$ in the codebook $\nu$.

**Theorem 3.2-sen-Poisson.** *If $\mu^n$ is a Poisson point process of intensity $\lambda_n$, then*

$$p_e(n) = \lambda_n \int_{\mathbb{R}^n} \mathbb{P}_0^n \left( D_0^n \in \mathcal{C}_x^n(\mu^n + \epsilon_x) \right) dx . \tag{49}$$

□

**Proof.** Using Slivnyak's theorem and the above notation, we have

$$\mathbb{E}_0^n \left( \left( \sum_{k \neq 0} \epsilon_{T_k^n + D_k^n} \right) (\mathcal{C}_0^n) \right) = \mathbb{E}^n \left( \left( \sum_k \epsilon_{T_k^n + D_k^n} \right) (\mathcal{C}_0^n(\mu^n + \epsilon_0)) \right),$$

with $\mu^n = \sum \epsilon_{T_k^n}$ a stationary Poisson point process. We now use Mecke's formula, which states that for all stationary point processes $\phi$ on $\mathbb{R}^n$ with intensity $\lambda$, defined on the probability space $(\Omega, \mathcal{F}, \mathbb{P}, \theta_x)$ and for all functions $v : \Omega_0 \times \mathbb{R}^n \to \mathbb{R}^+$, with $\Omega_0 = \Omega \cap \{\phi\{0\} = 1\}$,

$$\mathbb{E} \int_{\mathbb{R}^n} v(\theta_x(\omega), x) \phi(dx) = \lambda \mathbb{E}_\phi^0 \int_{\mathbb{R}^n} v(\omega, x) dx,$$

where $\mathbb{P}_\phi^0$ is the Palm probability of $\phi$. Applying this formula to the point process $\mu^n$ and to the function

$$v(\omega, x) = 1_{D_0 \in \mathcal{C}_{-x}(\mu^n(\omega) + \epsilon_{-x})}$$

gives:

$$\mathbb{E}^n \left( \left( \sum_k \epsilon_{T_k^n + D_k^n} \right) (\mathcal{C}_0^n(\mu^n + \epsilon_0)) \right) = \lambda_n \int_{\mathbb{R}^n} \mathbb{P}_0^n \left( D_0^n \in \mathcal{C}_{-x}^n(\mu^n + \epsilon_{-x}) \right) dx.$$

$\square$

## 7.2 Likelihood Ratio Representation under MLE

### 7.2.1 Poisson

**Theorem 3.1-sen-Poisson.** *Under the assumptions of Theorem 3-sen, if $\mu^n$ is Poisson, then*

$$p_e(n) \leq \lambda_n \int_{y^n \in \mathbb{R}^n} \mathbb{P}^n \left( \lambda_n W_\mathcal{D}^n \left( -\frac{1}{n} \ln(f^n(D^n)) \right) < 1, -\frac{1}{n} \ln \left( \frac{f^n(D^n - y^n)}{f^n(D^n)} \right) \leq 0 \right) dy^n$$

$$+ \mathbb{P}^n \left( \lambda_n W_\mathcal{D}^n \left( -\frac{1}{n} \ln(f^n(D^n)) \right) \geq 1 \right). \tag{50}$$

$\square$

**Proof.** From (28),

$$\begin{aligned} p_e(n) &\leq \int_{x^n \in \mathbb{R}^n} \mathbb{P}_0^n[(\mu^n - \epsilon_0)(\mathbb{F}(x^n)) > 0] f^n(x^n) dx^n \\ &\leq \int_{x^n \in \mathbb{R}^n} \min(1, \mathbb{E}_0^n[(\mu^n - \epsilon_0)(\mathbb{F}(x^n))]) f^n(x^n) dx^n, \end{aligned}$$

with $\mathbb{F}$ defined in (23). So, in the Poisson case, from Slivnyak's theorem,

$$\begin{aligned}
p_e(n) &\leq \int_{x^n \in \mathbb{R}^n} \min(1, \mathbb{E}^n[\mu^n(\mathbb{F}(x^n))])f^n(x^n)dx^n \\
&= \int_{x^n \in \mathbb{R}^n} \min(1, \lambda_n \text{Vol}(\mathbb{F}(x^n)))f^n(x^n)dx^n \\
&= \int_{x^n \in \mathbb{R}^n} \min\left(1, \lambda_n \int_{y^n \in \mathbb{R}^n} 1_{\mathbb{D}(x^n,y^n) \leq \mathbb{D}(x^n,0)} \, dy^n\right) f^n(x^n)dx^n \\
&= \lambda_n \int_{y^n \in \mathbb{R}^n} \int_{x^n \in \mathbb{R}^n} 1_{\lambda_n \text{Vol}(\mathbb{F}(x^n)) < 1} 1_{\mathbb{D}(x^n,y^n) \leq \mathbb{D}(x^n,0)} \, f^n(x^n)dx^n dy^n \\
&\quad + \int_{x^n \in \mathbb{R}^n} 1_{\lambda_n \text{Vol}(\mathbb{F}(x^n)) \geq 1} f^n(x^n)dx^n \\
&= \lambda_n \int_{y^n \in \mathbb{R}^n} \mathbb{P}^n(\lambda_n \text{Vol}(\mathbb{F}(D^n)) < 1, \mathbb{D}(D^n, y^n) \leq \mathbb{D}(D^n, 0))dy^n \\
&\quad + \mathbb{P}^n(\lambda_n \text{Vol}(\mathbb{F}(D^n)) \geq 1).
\end{aligned}$$

□

### 7.2.2 Matérn

The setting is that of Section 6.5. By the same arguments as those in the proofs of Theorems 3-sen-Matérn and 3.1-sen-Poisson, we get:

**Theorem 3.1-sen-Matérn.** *Under the assumptions of Theorem 3-sen-Matérn,*

$$\begin{aligned}
p_e(n) &\leq \lambda_n \int_{y^n \in \mathbb{R}^n} \mathbb{P}^n \left( \lambda_n \int_{z^n \in \mathbb{R}^n} 1_{-\frac{1}{n}\ln(f^n(z^n)) \geq \xi}, 1_{-\frac{1}{n}\ln\left(\frac{f^n(D^n - z^n)}{f^n(D^n)}\right) \leq 0} dz^n < 1, \right. \\
&\qquad \left. -\frac{1}{n}\ln\left(\frac{f^n(D^n - y^n)}{f^n(D^n)}\right) \leq 0 \right) 1_{\mathbb{D}(y^n,0) \geq \xi} dy^n \\
&\quad + \mathbb{P}^n \left( \lambda_n \int_{z^n \in \mathbb{R}^n} 1_{-\frac{1}{n}\ln(f^n(z^n)) \geq \xi}, 1_{-\frac{1}{n}\ln\left(\frac{f^n(D^n - z^n)}{f^n(D^n)}\right) \leq 0} dz^n \geq 1 \right). \quad (51)
\end{aligned}$$

□

## 7.3 Perturbation Analysis Representation

In this section, $n$ is fixed, and $\mathcal{C}_x^n(\nu) = \mathbb{C}^n(x, \nu)$ denotes the decoding region of $x$ in the codebook $\nu$. We make the assumption that the decoding regions satisfy the following monotonicity properties:

- for all $\mu^n$, all $z^n \in \mathbb{R}^d$ and all $y^n$ in the support of $\mu^n$, $\mathcal{C}_{y^n}^n(\mu^n + \epsilon_{z^n}) \subset \mathcal{C}_{y^n}^n(\mu^n)$.

- if $x^n \in \mathcal{C}^n_{y^n}(\mu^n)$ for some $y^n$ in the support of $\mu^n$, then for all $z^n$ such that $x^n \notin \mathcal{C}^n_{y^n}(\mu^n + \epsilon_{z^n})$, $x^n \in \mathcal{C}^n_{z^n}(\mu^n + \epsilon_{z^n})$.

These properties are for instance satisfied by the Voronoi and more generally the MLE regions.

For all $y^n \in \mathbb{R}^n$, consider the following random subset of $\mathbb{R}^n$, defined under the Palm probability of $\mu^n$ by:

$$\mathcal{G}^n_{y^n} = \{x^n \in \mathbb{R}^n : x^n \in C^n_0(\mu^n) \text{ and } x^n \in C^n_{y^n}(\mu^n + \epsilon_{y^n})\} . \quad (52)$$

In view of the above monotonicity assumption

$$\mathcal{G}^n_{y^n} = \{x^n \in \mathbb{R}^n : x^n \in C^n_0(\mu^n) \text{ and } x^n \notin C^n_0(\mu^n + \epsilon_{y^n})\} . \quad (53)$$

In the Voronoi cell case, $\mathcal{G}^n_{y^n}$ is the part of the Voronoi cell of 0 under $\mu^n$ which is "eaten out" by the addition of $y^n$ to $\mu^n$.

Similarly, for all $x^n \in \mathcal{C}^n_0$, let

$$\mathcal{H}^n_{x^n} = \{y^n \in \mathbb{R}^n : x^n \notin C^n_0(\mu^n + \epsilon_{y^n})\} . \quad (54)$$

In the Voronoi case, $\mathcal{H}^n_{x^n}$ is the part of the ball of center $x^n$ and of radius $|x^n|$ which is not included in $\mathcal{V}^n_0$. The union of the $\mathcal{H}^n_{x^n}$ domains when letting $x^n$ range over the whole Voronoi cell $\mathcal{V}^n_0$ is the so–called fundamental domain of $\mathcal{V}^n_0$ (see e.g. [16]).

**Theorem 3.3-sen-Poisson.** *Under the above assumptions, if $\mu^n$ is Poisson of intensity $\lambda = \lambda_n$, the error probability $p_e(n)$ associated with $\mathcal{C}^n$ and $D^n$ satisfies the relation*

$$\frac{d}{d\lambda} p_e(n) = \int_{\mathbb{R}^n} \int_{\mathbb{R}^n} \mathbb{P}^0_\lambda(x^n \in \mathcal{G}^n_{y^n}) f^n(x^n) dx^n dy^n$$
$$= \int_{\mathbb{R}^n} \mathbb{E}^0_\lambda(\mathit{Vol}(\mathcal{H}^n_{x^n})) f^n(x^n) dx^n , \quad (55)$$

*where $\mathbb{P}^0_\lambda$ is the Palm probability of the Poisson point process of intensity $\lambda$ on $\mathbb{R}^n$.* □

**Proof.** From the first order perturbation formula for functionals of marked Poisson point processes [3],

$$\frac{d}{d\lambda} \mathbb{P}^0_\lambda[D_0 \notin C_0(\mu_n)] = \int_{\mathbb{R}^n} \mathbb{E}^0_\lambda \left[ 1_{D_0 \notin C_0(\mu^n + \epsilon_{y^n})} - 1_{D_0 \notin C_0(\mu^n)} \right] dy^n .$$

From the monotonicity assumptions, the last quantity is equal to

$$\int_{\mathbb{R}^n} \mathbb{E}^0_\lambda \left[ 1_{D_0 \in C_{y^n}(\mu^n + \epsilon_{y^n})} 1_{D_0 \in C_0(\mu^n)} \right] dy^n$$

and hence to

$$\int_{\mathbb{R}^n} \mathbb{E}_\lambda^0 \left[ 1_{D_0 \notin C_0(\mu^n + \epsilon_{y^n})} 1_{D_0 \in C_0(\mu^n)} \right] dy^n \ .$$

The announced results then follow from independence and Fubini's theorem. □

This perturbation analysis method can be extended to higher order derivatives. These results might be of interest for the analysis of the so-called channel dispersion parameter (see [18]).

## 8 Error Exponents

### 8.1 Definition

Let $\mathcal{D}$ be some stationary ergodic additive noise. For all stationary and ergodic point processes $\mu^n$ of normalized logarithmic intensity $-h(\mathcal{D}) - \ln(\alpha)$ and all jointly stationary decoding regions $\mathcal{C}^n = \{\mathcal{C}_k^n\}_k$, let

$$p_e^{pp}(n, \mu^n, \mathcal{C}^n, \alpha, \mathcal{D}) \tag{56}$$

denote the probability of error associated with these data, as defined in (2). The $pp$ superscript is used to recall that the setting is the point process one described in Section 3. Let

$$p_{e,opt}^{pp}(n, \alpha, \mathcal{D}) \tag{57}$$

denote the infimum of $p_e^{pp}(n, \mu^n, \mathcal{C}^n, \alpha, \mathcal{D})$, where the infimum bears on all stationary and ergodic point processes $\mu^n$ with normalized logarithmic density bounded from below by $-h(\mathcal{D}) - \ln(\alpha)$ and on all jointly stationary decoding regions.

The error exponents for $(\alpha, \mathcal{D})$ are then defined as

$$\bar{\eta}(\alpha, \mathcal{D}) = \limsup_n -\frac{1}{n} \ln(p_{e,opt}^{pp}(n, \alpha, \mathcal{D})) \tag{58}$$

$$\underline{\eta}(\alpha, \mathcal{D}) = \liminf_n -\frac{1}{n} \ln(p_{e,opt}^{pp}(n, \alpha, \mathcal{D})). \tag{59}$$

Assuming these are identical, we denote this common limit by $\eta(\alpha, \mathcal{D})$.

For a fixed point process family $\mu = \{\mu^n\}_n$, with normalized logarithmic density $-h(\mathcal{D}) - \ln(\alpha)$ for all dimensions $n$, and for fixed decoding regions $\mathcal{C} = \{\mathcal{C}^n\}_n$, let

$$\pi(\mu, \mathcal{C}, \alpha, \mathcal{D}) = \lim_{n \to \infty} -\frac{1}{n} p_e^{pp}(n, \mu^n, \mathcal{C}^n, \alpha, \mathcal{D}) \ . \tag{60}$$

If the last limit does not exist, we define

$$\bar{\pi}(\mu, \mathcal{C}, \alpha, \mathcal{D}) = \limsup_{n \to \infty} -\frac{1}{n} \ln \left( p_e^{pp}(n, \mu^n, \mathcal{C}^n, \alpha, \mathcal{D}) \right) \tag{61}$$

$$\underline{\pi}(\mu, \mathcal{C}, \alpha, \mathcal{D}) = \liminf_{n \to \infty} -\frac{1}{n} \ln \left( p_e^{pp}(n, \mu^n, \mathcal{C}^n, \alpha, \mathcal{D}) \right) \ . \tag{62}$$

Since for all $\mu$ and $\mathcal{C}$ as above,
$$p_{e,opt}^{pp}(n, \alpha, \mathcal{D}) \leq p_e^{pp}(n, \mu^n, \mathcal{C}^n, \alpha, \mathcal{D}), \quad \forall n,$$
it follows that
$$\underline{\eta}(\alpha, \mathcal{D}) \geq \underline{\pi}(\mu, \mathcal{C}, \alpha, \mathcal{D}). \tag{63}$$

The aim of this section is to study the lower bounds $\underline{\pi}(\text{Poi}, \mathcal{L}(\mathcal{D}), \alpha, \mathcal{D})$ and $\underline{\pi}(\text{Mat}, \mathcal{L}(\mathcal{D}), \alpha, \mathcal{D})$ defined as follows: the decoding regions are the MLE cells $\mathcal{L}(\mathcal{D}) = \{\mathcal{L}^n(\mathcal{D})\}_n$, with $\mathcal{L}^n(\mathcal{D}) = \{\mathcal{L}_k^n(\mathcal{D})\}_k$ defined in (36) and the point processes are Poisson and Matérn respectively. Notice that the normalized logarithmic density fully characterizes the Poisson point process but not the Matérn. The $\underline{\pi}(\text{Mat}, \mathcal{L}(\mathcal{D}), \alpha, \mathcal{D})$ will be defined for a particular class of Matérn point processes satisfying the normalized logarithmic density constraint given above.

## 8.2 Poisson Point Process

### 8.2.1 White Gaussian Noise

Let $\mu$ be a sequence of Poisson processes $\mu^n$ of rates $\lambda_n = e^{nR}$ where $R = \frac{1}{2} \ln \frac{1}{2\pi e \alpha^2 \sigma^2}$ for $\alpha > 1$.

From (15), and because the MLE cells are equal to the Voronoi cells in the WGN displacement case, we have the following representation of the error probability:
$$\begin{aligned} p_e(n) &= \int_0^\infty \left(1 - e^{-\lambda_n V_B^n(r\sigma)}\right) g_1^n(r) dr \\ &= \int_0^\infty \left(1 - e^{-\lambda_n V_B^n(v\sigma\sqrt{n})}\right) g_1^n(v\sqrt{n})\sqrt{n} dv, \end{aligned} \tag{64}$$

**Theorem 4-wgn-Poisson.** *We have*
$$\underline{\pi}(\text{Poi}, \mathcal{L}(\text{WGN}), \alpha, \text{WGN}) = \begin{cases} \frac{\alpha^2}{2} - \frac{1}{2} - \ln \alpha & \text{if } 1 \leq \alpha < \sqrt{2} \\ \frac{1}{2} - \ln 2 + \ln \alpha & \text{if } \sqrt{2} \leq \alpha < \infty \end{cases}. \tag{65}$$
□

**Proof.** This theorem is a direct corollary of Theorem 4-sen below. We nevertheless give the main ideas of a direct proof here and details on this direct proof in Appendix 10.3.

We have
$$g_1^n(v\sqrt{n}) = e^{-n\left(\frac{v^2}{2} - \frac{1}{2} - \ln(v) + o(1)\right)},$$
see equation (129) and the discussion of the appendix, and
$$\mathbb{P}^n(\mu^n(B^n(v\sigma\sqrt{n}\vec{u}, 0)) > 0) = 1 - e^{-\lambda_n V_B^n(v\sigma\sqrt{n})} = e^{-n\left((\ln\alpha - \ln v)^+ + o(1)\right)}.$$

We may therefore write the probability of error in (64) as
$$p_e(n) = \int_0^\infty e^{-n\left(\frac{v^2}{2} - \frac{1}{2} - \ln v + (\ln\alpha - \ln v)^+ + o(1)\right)} dv.$$

We observe that the exponent of the integrand at $v$ is of the form $a(v) + b(v)$, where

$$a(v) = \frac{v^2}{2} - \frac{1}{2} - \ln(v) \qquad (66)$$

pertains to the additive noise and

$$b(v) = (\ln \alpha - \ln v)^+ \qquad (67)$$

to the ball having some point. Carrying out the optimization:

$$\text{Minimize } a(v) + b(v) \text{ over } v \geq 0, \qquad (68)$$

gives $\frac{\alpha^2}{2} - \frac{1}{2} - \ln \alpha$ when $1 < \alpha < \sqrt{2}$ and $\frac{1}{2} - \ln 2 + \ln \alpha$ when $\alpha > \sqrt{2}$. □

### 8.2.2 Beyond White Gaussian Noise

The setting of this subsection is that of Section 6.3. From (30),

$$p_e(n) \leq \int_{u \in \mathbb{R}} (1 - \exp(-\lambda_n W_\mathcal{D}^n(u))) \, \rho_\mathcal{D}^n(du) \,. \qquad (69)$$

Below, in order to analyze error exponents, we proceed in three steps:

1. We give sufficient conditions for the entropy spectrum $\rho_\mathcal{D}^n(.)$ to satisfy an LDP;

2. We give a log scale asymptotic for the stun level volume $W_\mathcal{D}^n(.)$;

3. We use the Laplace–Varadhan integral lemma to deduce the error exponent from the LDP satisfied by $\rho_\mathcal{D}^n(.)$.

The definition retained below for an LDP is that on p. 5 of [7], which features a probability space $(\mathcal{X}, \mathcal{B})$, where $\mathcal{X}$ is a topological space and where $\mathcal{B}$ is a $\sigma$-field which may be thinner than the Borel $\sigma$-field of $\mathcal{X}$.

**1. LDP for the entropy spectrum.**

**1.a. Stationary ergodic noise.**

The key assumption under which the stationary ergodic case will be analyzed in what follows is summarized below:

**H-sen**

1. For all $n$, $D^n = (\mathcal{D}_1, \ldots, \mathcal{D}_n)$ admits a density $f^n$ on $\mathbb{R}^n$;

2. For all $n$, the differential entropy of $D^n$, $h(D^n)$, is well defined;

3. The differential entropy rate of $\{\mathcal{D}_k\}$,

$$\begin{aligned} h(\mathcal{D}) &= \lim_{n\to\infty} \frac{1}{n} h(D^n) \\ &= \lim_{n\to\infty} -\frac{1}{n} \mathbb{E}\left(\ln(f^n(D^n))\right) \\ &= \lim_{n\to\infty} -\frac{1}{n} \int_{\mathbb{R}^n} \ln(f^n(x^n)) f^n(x^n) dx^n \\ &= \lim_{n\to\infty} \int_{\mathbb{R}} u \rho_{\mathcal{D}}^n(du), \end{aligned}$$

exists and is finite;

4. The entropy spectrum $\rho_{\mathcal{D}}^n(.)$, namely the law of the random variables $\{-\frac{1}{n}\ln(f^n(D^n))\}$, satisfies an LDP (on the real line endowed with its Borel $\sigma$-field), with good (in particular lower semicontinuous) and convex[1] rate function $I(x)$.

A simple sufficient condition for 4. to hold is that the conditions of the Gärtner-Ellis Theorems hold, namely that the limit

$$\lim_{n\to\infty} \frac{1}{n} \ln\left(\mathbb{E}\left((f^n(D^n))^{-\theta}\right)\right) = G(\theta) \tag{70}$$

exists as an extended real number, is finite in some neighborhood of the origin, and is essentially smooth (see Definition 2.3.5 in [7]). From the Gärtner-Ellis Theorem, the family of measures $\rho_{\mathcal{D}}^n(.)$ then satisfies an LDP with good and convex rate function

$$I(x) = \sup_{\theta} \left(\theta x - G(\theta)\right). \tag{71}$$

**Remark 6.** In order to analyze error exponents, there are other options than assuming that the entropy spectrum satisfies an LDP. In certain particular cases (see below), it is more convenient to assume that certain empirical measures of the noise process satisfy an LDP to deduce an LDP for $\rho_{\mathcal{D}}^n(.)$ when using the contraction principle [7]. We will come back to this in what follows.

**1.b. White noise.**

In the particular case where the noise has i.i.d. components with density $f$, we deduce from Cramér's theorem that

$$I(x) = \sup_{\theta} \left(\theta x - \ln\left(\mathbb{E}\left(f(D)^{-\theta}\right)\right)\right),$$

with $D$ a random variable with density $f$. A direct calculation gives that $\rho_{\mathcal{D}}^n(.)$ satisfies an LDP with rate function

$$I(x) = \theta(x)x - \ln\left(\mathbb{E}\left(f(D)^{-\theta(x)}\right)\right), \tag{72}$$

---

[1] Some of the results derived below do not require this convexity assumption.

where $\theta(x)$ is the unique solution of the equation $x = \phi(\theta)$ with

$$\phi(\theta) = \frac{\mathbb{E}\left((d(D))^{-\theta}\ln(1/f(D))\right)}{\mathbb{E}\left((f(D))^{-\theta}\right)}.$$

The uniqueness follows from the fact that this function is monotonic in $\theta$ (as easily checked by differentiating and using the Cauchy-Schwartz inequality).

Notice that the rate function $I(.)$ is not necessarily a good rate function. A sufficient condition - assumed in what follows - is that 0 is in the interior of the set

$$\{\theta \;:\; \mathbb{E}\left((f(D))^{-\theta}\right) < \infty\}$$

(see [7], Lemma 2.2.20).

Here are a few examples.

- **WGN.** We deduce from the results of §10.1 that

$$\mathbb{E}\left(f(D)^{-\theta}\right) = (2\pi\sigma^2)^{\theta/2}\sqrt{\frac{1}{1-\theta}}$$

  when $\theta < 1$ and is equal to $\infty$ for $\theta > 1$. So

$$\begin{aligned} I(u) &= \sup_{\theta}\left(\theta u - \theta/2\ln(2\pi\sigma^2) + 1/2\ln(1-\theta)\right) \\ &= u - h(D) - \frac{1}{2}\ln(2u - \ln(2\pi\sigma^2))\;. \end{aligned}$$

  Hence

$$I(u) = \begin{cases} +\infty & \text{for } u \leq \frac{1}{2}\ln(2\pi\sigma^2); \\ u - \frac{1}{2}\ln(2e\pi\sigma^2) - \frac{1}{2}\ln(2u - \ln(2\pi\sigma^2)) & \text{otherwise,} \end{cases} \quad (73)$$

  which is a good and convex rate function.

- **White symmetric exponential noise.** Here

$$\mathbb{E}\left(f(D)^{-\theta}\right) = (\sqrt{2}\sigma)^{\theta}\mathbb{E}\left(\exp\left(\theta\frac{|D|\sqrt{2}}{\sigma}\right)\right) = (\sqrt{2}\sigma)^{\theta}\frac{1}{1-\theta}, \quad \theta < 1\;.$$

  So

$$I(u) = \sup_{\theta}\left(\theta u - \theta\ln(\sqrt{2}\sigma) + \ln(1-\theta)\right),$$

  that is

$$I(u) = \begin{cases} +\infty & \text{for } u \leq \ln(\sqrt{2}\sigma); \\ u - h(D) - \ln(u - \ln(\sqrt{2}\sigma)) & \text{otherwise,} \end{cases} \quad (74)$$

  which is a good and convex rate function.

- **White uniform noise.** Here

$$\mathbb{E}\left(f(D)^{-\theta}\right) = (2\sqrt{3}\sigma)^{\theta},$$

so that $G(\theta) = \theta \ln(2\sqrt{3}\sigma)$ and

$$I(u) = \begin{cases} \infty & \text{if } u \neq \ln(2\sqrt{3}\sigma) \\ 0 & \text{if } u = \ln(2\sqrt{3}\sigma), \end{cases} \tag{75}$$

which is a good and convex rate function.

**Large Deviations on Empirical Measures** Let us now illustrate Remark 6 by a few observations on the fact that the LDP on the entropy spectrum may follow from an LDP on empirical measures, namely from Sanov's theorem since this section focuses on the white noise case. The interest of this observation is not purely theoretical as this derivation will also provide a new representation of the rate function $I(x)$ as the solution of an optimization problem based on relative entropy as we shall see.

Below, we will denote by $S$ the support of $f$, namely the closure of $\{x \in \mathbb{R} : f(x) > 0\}$. Let $K(\tau \| \phi)$ be the relative entropy (or Kullback-Leibler divergence) of the probability law $\tau(dx)$ w.r.t. the probability law $\phi(dx) = f(x)dx$, defined as follows: if $\tau$ is absolutely continuous w.r.t. $\phi$, then

$$K(\tau \| \phi) = \int_{\mathbb{R}} \ln\left(r(x)\right) r(x) f(x) dx, \tag{76}$$

with $r = \frac{d\tau}{d\phi}$. If $\tau$ does not satisfy the above property, then $K(\tau \| \phi) = \infty$.

Note that $\tau$ is absolutely continuous w.r.t. $\phi$ iff it admits a density $g$ such that $g(y) = 0$ when $f(y) = 0$ for a.a. $y$. In this case,

$$K(\tau \| \phi) = \int_S \ln\left(\frac{g(x)}{f(x)}\right) g(x) dx.$$

With an abuse of notation we will continue to use the notation $\phi$ for its restriction to $S$. So we now have $\phi \in \mathbb{M}_1(S)$, with $\mathbb{M}_1(S)$ the topological space of probability measures on $S$ endowed with the topology of weak convergence. Let $\mathcal{M}_1(S)$ denote the Borel $\sigma$-algebra corresponding to this topology. The empirical measures

$$\nu^n = -\frac{1}{n} \sum_{i=1}^n \epsilon_{\mathcal{D}_i}$$

are $\mathbb{M}_1(S)$-valued random variables. From Sanov's theorem (see e.g. [7], Theorem 6.2.10), the laws of the random measures $\nu^n$ satisfy an LDP on the Hausdorff space $(\mathbb{M}_1(S), \mathcal{M}_1(S))$, with good and convex rate function $K(. \| \phi)$.

**Lemma 1-wn.** *Assume that the function $x \to \ln(f(x))$ from $S$ to $\mathbb{R}$ is continuous and bounded. Then the family of measures $\rho_{\mathcal{D}}^n$ on $\mathbb{R}$ satisfies an LDP with good and convex rate function*

$$I(u) = \inf_{\tau \in \mathbb{M}_1(S):\ -\int_S \ln(f(x))\tau(dx)=u} K(\tau||\phi) \qquad (77)$$

$$= u - \sup_{\tau \in \mathbb{M}_1(S):\ K(\tau||\phi)+h(\tau)=u} h(\tau). \qquad (78)$$

□

**Proof.** Under the above assumption, the mapping

$$\tau \to \int_S \ln(f(x))\tau(dx)$$

is continuous from $(\mathbb{M}_1(S), \mathcal{M}_1(S))$ to $(\mathbb{R}, \mathcal{B}(\mathbb{R}))$. The contraction principle shows that the real-valued random variables

$$-\frac{1}{n}\sum_{i=1}^n \ln(f(D_i)) = \int_S l(x)\nu^n(dx),$$

with $l(x) = -\ln(f(x))$, satisfy an LDP with rate function (77). The infimum in (77) is reached on the set of laws absolutely continuous w.r.t. $f(x)dx$, namely the laws with a density $g$ such that $g = 0$ if $f = 0$. Hence

$$I(u) = \inf_{g:\ f=0 \Rightarrow g=0,\ \text{and}\ -\int_S \ln(f(x))g(x)dx=u} K(\tau||f).$$

It is shown in Appendix 10.5 that for all $g$ absolutely continuous w.r.t. $f(x)dx$ and with a finite relative entropy w.r.t. $f$, the differential entropy of $g$ exists. Hence the following equality makes sense:

$$\int_S \ln(f(x))g(x)dx = \int_S -\ln\left(\frac{g(x)}{f(x)}\right)g(x)dx + \int_S \ln(g(x))g(x)dx$$
$$= -K(\tau||\phi) - h(\tau),$$

so that

$$I(u) = \inf_{\tau \in \mathbb{M}_1(S):\ K(\tau||\phi)+h(\tau)=u} K(\tau||\phi) = u - \sup_{\tau \in \mathbb{M}_1(S):\ K(\tau||\phi)+h(\tau)=u} h(\tau).$$

□

**Remark 7.** Notice that the last lemma requires densities $f$ which are 1) bounded from above and from below by positive constants, and hence with a bounded support (hence the WGN case is not covered); 2) continuous on $S$. Stronger versions of the contraction principle like those considered in [9] might be useful for extending Lemma 1-wn to e.g. the case of densities with unbounded support. The extension should take the following observation into account: as shown in Appendix 10.6, there exist densities $f$ and $g$ with infinite support and such that 1) $f$ and $g$ have the same support, 2) $f$ has finite differential entropy, 3) the relative entropy of $g(x)dx$ w.r.t. $f(x)dx$ is finite and 4) the differential entropy of $g$ is not defined (i.e. has a positive part and a negative part which are both infinite. This shows that (77)–(78) cannot be extended as such to the case where one replaces the boundedness of $\ln(f(x))$ on $S$ by the weaker assumption that $f$ has a well defined differential entropy.

**1.c. Beyond white noise.**

- **Colored Gaussian noise.** In the colored Gaussian noise case, we deduce from the Grenander–Szegö Theorem that the Gärtner-Ellis Theorem holds with

$$G(\theta) = \frac{\theta}{2}\ln(2\pi) - \frac{1}{2}\ln(1-\theta) + \frac{\theta}{2}\ln\left(\frac{1}{2\pi}\int_{-\pi}^{\pi}\ln(g(\beta))d\beta\right) ,$$

when $\theta < 1$ and $G(\theta) = \infty$ for $\theta > 1$. So, by the same arguments as above,

$$I(u) = \begin{cases} \infty & \text{if } u \leq \frac{1}{4\pi}\int_{-\pi}^{\pi}\ln(2\pi g(\beta))d\beta; \\ u - h(\mathcal{D}) - \frac{1}{2}\ln\left(2u - \frac{1}{2\pi}\int_{-\pi}^{\pi}\ln(2\pi g(\beta))d\beta\right) & \text{otherwise,} \end{cases} \quad (79)$$

with

$$h(\mathcal{D}) = \frac{1}{4\pi}\int_{-\pi}^{\pi}\ln(2\pi e g(\beta))d\beta ,$$

which is a good, convex and continuous rate function.

- **Markov noise.** We continue the thread of Remark 6 on the derivation of the LDP on the entropy spectrum from an LDP on empirical measures. We let $S \subseteq \mathbb{R}^2$ denote the support of the measure on $\mathbb{R}^2$ with density $a(x)p(y|x)$ with respect to Lebesgue measure, i.e. the closure of $\{(x,y) : a(x)p(y|x) > 0\}$. We denote by $\mathbb{M}_1(S)$ the Hausdorff space of probability measures on $S$. The topology on $\mathbb{M}_1(S)$ is again that of weak convergence and we denote by $\mathcal{M}_1(S)$ the Borel $\sigma$-field for this topology. Finally, let $\mathrm{SM}_1(S)$ be the set of probability measures of $\mathbb{M}_1(S)$ of the form $b(x)q(y|x)$ where $y \mapsto q(y|x)$ is a probability density on the real line for each $x \in \mathbb{R}$, $x \mapsto b(x)$ is a probability density on $\mathbb{R}$, and

$$\int b(x)q(y|x)dx = b(y) .$$

We assume that the Markov chain satisfies the uniformity condition U of Section 6.3 in [7]. Then Corollary 6.5.10 in [7] shows that the empirical measures

$$\frac{1}{n}\sum_{i=1}^{n}\epsilon_{\mathcal{D}_i,\mathcal{D}_{i+1}}$$

satisfy an LDP on the measurable space $(\mathbb{M}_1(S), \mathcal{M}_1(S))$ with the good and convex rate function

$$I(\tau) = \begin{cases} K(\tau || \tau_1 \otimes P) & \text{if } \tau \in \text{SM}_1(S) \\ \infty & \text{otherwise.} \end{cases} \quad (80)$$

Here $K$ is the Kullback-Leibler divergence of measures on $\mathbb{R}^2$, $\tau_1$ and $\tau_2$ are the marginals of $\tau$, and $\tau_1 \otimes P$ is the measure $\tau_1(dx)P(x,y)dy \in \mathbb{M}_1(\mathbb{R}^2)$.

**Lemma 1-mn.** *Under the above assumptions, if the mapping $(x, y) \to \ln(p(y \mid x))$ from $S$ to $\mathbb{R}$ is continuous and bounded, then the family of measures $\rho_\mathcal{D}^n(.)$ satisfies an LDP with the good and convex rate function*

$$I(u) = \inf_{\tau \in \text{SM}_1(S): \int_{\pi(x)p(y|x)>0} \ln(p(y|x))\tau(dxdy) = u} K(\tau || \tau_1 \otimes P) \quad (81)$$

$$= u - \sup_{\tau \in \text{SM}_1(S): \ K(\tau || \tau_1 \otimes P) + h(\tau_2 | \tau_1) = u} h(\tau_2 | \tau_1), \quad (82)$$

*provided the last function is convex and admits an essentially smooth Fenchel-Legendre transform (see Defn. 2.3.5 of [7]).* □

**Proof.** Under the above assumptions, the mapping

$$\tau \to \int_S \ln(p(y \mid x))\tau(dxdy)$$

is continuous from $(\mathbb{M}_1(S), \mathcal{M}_1(S))$ to $(\mathbb{R}, \mathcal{B}(\mathbb{R}))$. The contraction principle shows that the random variables

$$A_n = -\frac{1}{n} \sum_{i=1}^n \ln(p(\mathcal{D}_{i+1} \mid \mathcal{D}_i))$$

satisfy an LDP on $\mathbb{R}$ with good and convex rate function (81). Let

$$X_n = -\frac{1}{n} \ln(\pi(\mathcal{D}_1)) - \frac{1}{n} \sum_{i=1}^n \ln(P(\mathcal{D}_i, \mathcal{D}_{i+1}))$$

and

$$B_n = -\frac{1}{n} \ln(\pi(\mathcal{D}_1)) - \frac{1}{n} \sum_{i=1}^n \ln(P(\mathcal{D}_i, \mathcal{D}_{i+1})) + \frac{1}{n} \ln(\pi(\mathcal{D}_{n+1})).$$

The random variables $B_n$ satisfy an LDP with the same good rate function $I(u)$. This follows from the application of the contraction principle to the mapping

$$\tau \to \int_S \left( -\ln(p(y \mid x)) - \ln(\pi(x)) + \ln(\pi(y)) \right) \tau(dxdy) \ .$$

We also have $B_n \leq X_n \leq A_n$ for all $n$. It then follows that $X_n$ satisfies an LDP with good and convex rate function $I$ (see Appendix 10.4).

The optimization problem (82) can be rewritten under the following form:

$$I(u) = \inf_{\tau \in \mathrm{SM}_1(S):\ K(\tau||\tau_1 \otimes P) + h(\tau_2|\tau_1) = u} K(\tau||\tau_1 \otimes P)$$
$$= u - \sup_{\tau \in \mathrm{SM}_1(S):\ K(\tau||\tau_1 \otimes P) + h(\tau_2|\tau_1) = u} h(\tau_2|\tau_1).$$

because $h(\tau_2|\tau_1)$ is well defined on the set where the optimum is achieved (see the second part of Appendix 10.5). □

## 2. Log scale asymptotics of the stun level volumes.

### 2.a Stationary ergodic noise.

**Lemma 2-sen.** *Assume that the H-sen assumptions hold. Then*

$$\sup_{s<u}(s - I(s)) \leq \liminf_{n \to \infty} \frac{1}{n} \ln(W_{\mathcal{D}}^n(u)) \leq \limsup_{n \to \infty} \frac{1}{n} \ln(W_{\mathcal{D}}^n(u)) \leq \sup_{s \leq u}(s - I(s)). \tag{83}$$

*The function*

$$J(u) = \sup_{s \leq u}(s - I(s)), \tag{84}$$

*which will be referred to as the* volume exponent, *is upper semicontinuous.* □

**Proof.** From (35),

$$W_{\mathcal{D}}^n(u) \geq \int e^{n\phi(s)} \rho_{\mathcal{D}}^n(ds),$$

with

$$\phi(s) = s 1_{s<u} - \infty 1_{s \geq u}.$$

with $\rho_{\mathcal{D}}^n$ satisfying an LDP and since the function $\phi$ is lower semicontinuous, the lower bound is proved as in Lemma 4.3.4 in [7]. Similarly, the function

$$\widetilde{\phi}(s) = s 1_{s \leq u} - \infty 1_{s > u}$$

is upper semicontinuous and the upper bound is proved as in Lemma 4.3.6 in [7]. In both cases, it should be noticed that the proofs in [7] actually allow for functions $\phi$ with values in $\{-\infty\} \cup \mathbb{R}$.

Let us now show that the upper semicontinuity of the function $g(s) = s - I(s)$ implies that of the function $J(u) = \sup_{s \leq u} g(s)$. We have to show that

$$J(u) \geq \lim_{\epsilon \to 0} \sup_{s \in [u-\epsilon, u+\epsilon]} J(s) = \lim_{\epsilon \to 0} J(u + \epsilon), \tag{85}$$

where the last inequality follows from the fact that $J$ is non-decreasing. Hence, using monotonicity again, we have to show that $J$ is right-continuous. We have

$$J(u+\epsilon) = J(u) + \sup_{s \in [u,u+\epsilon]} (g(s) - J(u))^+ ,$$

with $a^+ = \max(a, 0)$. So, either $g(s) \leq J(u)$ for all $s \in [u, u+\epsilon]$, in which case $J(u+\epsilon) = J(u)$ and the right-continuity is trivially satisfied, or $g(s) > J(u)$ for some $s \in [u, u+\epsilon]$, in which case

$$J(u+\epsilon) = \sup_{[u,u+\epsilon]} g(s).$$

It then follows from the upper semicontinuity of the function $g(s)$ that

$$J(u) \geq g(u) \geq \lim_{\epsilon \to 0} \sup_{[u,u+\epsilon]} g(s) = \lim_{\epsilon \to 0} J(u+\epsilon),$$

so that (85) and hence right-continuity hold in this case too. $\square$

Here are two observations:

- Since $I(h(\mathcal{D})) = 0$, it follows from (84) that $J(h(\mathcal{D})) \geq h(\mathcal{D})$. The concavity of the function $x \to x - I(x)$ implies that this function is non decreasing on the interval $(-\infty, h(\mathcal{D})]$. Hence, from (84),

$$J(h(\mathcal{D})) = h(\mathcal{D}) . \tag{86}$$

- At all points $u$ of continuity of $J$

$$\lim_{n \to \infty} \frac{1}{n} \ln(W_{\mathcal{D}}^n(u)) = J(u) . \tag{87}$$

### 2.b White noise.

Below, we give a few basic examples and we use the following observation: consider the white noise case with law $\phi$ admitting the density $f$ on $\mathbb{R}$. When (78) holds, it follows from Lemma 2-sen that

$$\sup_{\tau: \ K(\tau||\phi)+h(\tau)<u} h(\tau) \leq \liminf_{n \to \infty} \tfrac{1}{n} \ln(W_{\mathcal{D}}^n(u)) \tag{88}$$

$$\leq \limsup_{n \to \infty} \tfrac{1}{n} \ln(W_{\mathcal{D}}^n(u)) \leq \sup_{\tau: \ K(\tau||\phi)+h(\tau)\leq u} h(\tau) .$$

- **WGN.** In the white Gaussian noise case, it immediately follows from Lemma 2-sen that

$$J(u) = \begin{cases} -\infty & \text{for } u \leq \tfrac{1}{2}\ln(2\pi\sigma^2) \\ \tfrac{1}{2}\ln(2\pi e\sigma^2) + \tfrac{1}{2}\ln\left(2u - \ln(2\pi\sigma^2)\right) & \text{otherwise .} \end{cases} \tag{89}$$

There is of course a more direct derivation of this result based on the log scale asymptotic of the volume of the ball of radius $\sqrt{n2\sigma^2(u - 1/2\ln(2\pi\sigma^2))}$.

Here is a third (heuristic) way: in spite of the fact that there is no justification of (78) in the WGN case, it is easy to see that the optimization problem (88) gives the right answer too. We have

$$-\ln(f(x)) = \frac{1}{2}\ln(2\pi\sigma^2) + \frac{x^2}{2\sigma^2}$$

and (88) boils down to finding the distribution function $\tau$ of maximal differential entropy and with variance less than $u - \frac{1}{2}\ln(2\pi\sigma^2)$. This of course requires that $u \geq \frac{1}{2}\ln(2\pi\sigma^2)$. The solution to this optimization problem is the Gaussian distribution function $\tau_u^*$ with variance $2\sigma^2(u - \frac{1}{2}\ln(2\pi\sigma^2))$. Hence

$$\lim_{n\to\infty} \frac{1}{n} \ln(W_{\mathcal{D}}^n(u)) = h(\tau_u^*) = \frac{1}{2}\ln(2\pi e) + \frac{1}{2}\ln\left(2\sigma^2(u - \frac{1}{2}\ln(2\pi\sigma^2))\right)$$

$$= \frac{1}{2}\ln(2\pi e\sigma^2) + \frac{1}{2}\ln\left(2u - \ln(2\pi\sigma^2)\right).$$

- **White symmetric exponential noise.** We can again evaluate the volume by three different ways. The first is Lemma 2-sen and gives

$$J(u) = \begin{cases} -\infty & \text{for } u \leq \ln(\sqrt{2}\sigma) \\ \ln(\sqrt{2}e\sigma(u - \ln(\sqrt{2}\sigma))) & \text{otherwise .} \end{cases} \quad (90)$$

The second is geometric: using the fact that the volume of the $L_1$ ball of radius $a$ is $\frac{(2a)^n}{n!}$, we get that for $u > \ln(\sqrt{2}\sigma)$,

$$W_{\mathcal{D}}^n(u) = \text{Vol}\left\{y^n \text{ s.t. } \frac{1}{n}\sum |y_i|\frac{\sqrt{2}}{\sigma} < u - \ln(\sqrt{2}\sigma)\right\}$$

$$= \frac{(\sqrt{2}\sigma(u - \ln(\sqrt{2}\sigma))n)^n}{n!}$$

$$= \exp\left(n\ln(\sqrt{2}e\sigma(u - \ln(\sqrt{2}\sigma))) + o(n)\right).$$

The third (heuristic) way is based on the solution of the optimization problem (88). The latter reduces to finding the distribution with maximal entropy on $\mathbb{R}$ with mean absolute value equal to $t = \sigma(u - \ln(\sqrt{2}\sigma))/\sqrt{2}$. This is the symmetric exponential with parameter $\sigma' = t\sqrt{2}$ (see Appendix 10.2) which has for differential entropy:

$$\ln(\sqrt{2}e\sigma') = \ln(\sqrt{2}e\sigma(u - \ln(\sqrt{2}\sigma))).$$

The three approaches are hence consistent.

- **White uniform noise.** From Lemma 2-sen,

$$J(u) = \begin{cases} -\infty & \text{for } u < \ln(2\sqrt{3}\sigma) \\ \ln(2\sqrt{3}\sigma) & \text{for } u \geq \ln(2\sqrt{3}\sigma) \end{cases}, \quad (91)$$

which also fits with the other viewpoints.

In the first two examples, the function $u \to J(u)$ is continuous. In the third one, the function is only upper semicontinuous.

**2.c Beyond white noise.**

- **Colored Gaussian noise.** From (79) and (84), in the colored noise case

$$J(u) = \begin{cases} -\infty & \text{if } u \leq \frac{1}{4\pi}\int_{-\pi}^{\pi} \ln(2\pi g(\beta))d\beta; \\ \frac{1}{4\pi}\int_{-\pi}^{\pi} \ln(2\pi e g(\beta))d\beta + \frac{1}{2}\ln\left(2u - \frac{1}{2\pi}\int_{-\pi}^{\pi} \ln(2\pi g(\beta))d\beta\right) & \\ & \text{otherwise}. \end{cases} \quad (92)$$

This function is continuous.

- **Markovian noise.**

Under the assumptions of Lemma 1-mn,

$$J(u) = \sup_{\tau \in \text{SM}_1(\mathbb{R}^2):\ K(\tau||\tau_1 \otimes P) + h(\tau_2|\tau_1) \leq u} h(\tau_2|\tau_1) . \quad (93)$$

**3. Error exponents.**

**3.a. Stationary ergodic noise.**

**Theorem 4-sen-Poisson.** *Assume that $\mu^n$ is Poisson with normalized logarithmic intensity $-h(\mathcal{D}) - \ln(\alpha)$ with $\alpha > 1$ and that decoding is MLE. Assume the noise $\mathcal{D}$ to be stationary and ergodic and such that the Assumptions H-sen hold. Then the associated error exponent is such that*

$$\underline{\pi}(Poi, \mathcal{L}(\mathcal{D}), \alpha, \mathcal{D}) \geq \inf_u \{F(u) + I(u)\} , \quad (94)$$

*where $I(u)$ is the rate function of $\rho_\mathcal{D}^n$ (defined in (71)) and*

$$F(u) = (\ln(\alpha) + h(\mathcal{D}) - J(u))^+ , \quad (95)$$

*where $J(u) = \sup_{s \leq u}(s - I(s))$ is the volume exponent defined in Lemma 2-sen.*
□

**Proof.** We use (69) and the bound
$$1 - e^{-\lambda_n W_{\mathcal{D}}^n(u)} \leq \min(1, \lambda_n W_{\mathcal{D}}^n(u))$$
to write
$$p_e(n) = \int_u e^{-n\phi_n(u)} \rho_{\mathcal{D}}^n(du),$$
with
$$\phi_n(u) = \left(\ln(\alpha) + h(\mathcal{D}) - \frac{1}{n}\ln(W_{\mathcal{D}}^n(u))\right)^+.$$

In order to conclude, we use Theorem 2.3 in [20]. Since the law $\rho_{\mathcal{D}}^n(du)$ satisfies an LDP with good rate function $I(u)$, it is enough to prove that for all $\delta > 0$, there exists $\epsilon > 0$ such that

$$\liminf_{n \to \infty} \inf_{(u-\epsilon, u+\epsilon)} \left(\ln(\alpha) + h(\mathcal{D}) - \frac{1}{n}\ln(W_{\mathcal{D}}^n(u))\right)^+ \geq (\ln(\alpha) + h(\mathcal{D}) - J(u)))^+ - \delta.$$

Since the function $u \to W_{\mathcal{D}}^n(u)$ is non decreasing, the last relation is equivalent to

$$\liminf_{n \to \infty} \left(\ln(\alpha) + h(\mathcal{D}) - \frac{1}{n}\ln(W_{\mathcal{D}}^n(u+\epsilon))\right)^+ \geq (\ln(\alpha) + h(\mathcal{D}) - J(u)))^+ - \delta.$$

Hence it is enough to show that for all $\delta > 0$, there exists $\epsilon > 0$ such that

$$\lim_{n \to \infty} \left(\ln(\alpha) + h(\mathcal{D}) - \sup_{m \geq n} \frac{1}{m}\ln(W_{\mathcal{D}}^m(u+\epsilon))\right)^+ \geq (\ln(\alpha) + h(\mathcal{D}) - J(u)))^+ - \delta.$$

There are two cases: if $\ln(\alpha) + h(\mathcal{D}) - J(u) \leq 0$, the result is obvious. if $\ln(\alpha) + h(\mathcal{D}) - J(u) > 0$, then we have to prove that for all $\delta$, there exists an $\epsilon$ such that
$$\lim_{n \to \infty} \sup_{m \geq n} \frac{1}{m}\ln(W_{\mathcal{D}}^m(u+\epsilon)) \leq \sup_{s \leq u}(s - I(s)) + \delta$$

But from Lemma 2-sen,
$$\lim_{n \to \infty} \sup_{m \geq n} \frac{1}{m}\ln(W_{\mathcal{D}}^m(u+\epsilon)) \leq \sup_{s \leq u+\epsilon}(s - I(s)).$$

Hence it is enough to show that for all $\delta$, there exists an $\epsilon$ such that
$$\sup_{s \leq u}(s - I(s)) \geq \sup_{s \leq u+\epsilon}(s - I(s)) - \delta.$$

This follows from the fact that the function $J(u)$ is upper semicontinuous. □

Notice that all terms in the final expression to be minimized, namely
$$(\ln(\alpha) + h(\mathcal{D}) - J(u))^+ + I(u)$$
can be traced back as follows:

- $e^{-(\ln(\alpha)+h(\mathcal{D}))}$ is the intensity $\lambda_n$;
- $e^{nJ(u)}$ is the volume of the stun level set for level $u$;
- $e^{-nI(u)}$ is the value of the density of the entropy spectrum at $u$;
- the positive part stems from the minimum of the mean number of points in the above set and 1.

**3.b. White noise.** Here are direct applications of the last results.

- **WGN.** In the white Gaussian noise case, we find back the minimization problem (68) and hence the result of Theorem 4-wn-Poisson from (94) as follows: use the formula (73) for $I$ and the formula (89) for $J$ in (94) and pose $v = \sqrt{2(u - \frac{1}{2}\ln(2\pi\sigma^2))}$.

- **White symmetric exponential noise.** We get from the formula (74) for $I$ and the formula (90) for $J$ that, in this case, the function to minimize in (94) is

$$v - 1 - \ln(v) + (\ln(\alpha) - \ln(v))^+,$$

for $v > 0$. So in this case,

$$\underline{\pi}(\text{Poi}, \mathcal{L}(\mathcal{D}), \alpha, \mathcal{D}) \geq \begin{cases} \alpha - 1 - \ln\alpha & \text{if } 1 \leq \alpha < 2 \\ 1 - 2\ln 2 + \ln\alpha & \text{if } 2 \leq \alpha. \end{cases} \quad (96)$$

- **White uniform noise.** Here we deduce from (91) and (75) and (94) that

$$\underline{\pi}(\text{Poi}, \mathcal{L}(\mathcal{D}), \alpha, \mathcal{D}) \geq F(\ln(2\sqrt{3}\sigma)) = \ln(\alpha) . \quad (97)$$

**3.c Beyond white noise.**

- **Colored Gaussian noise.** In the colored Gaussian noise case, Theorem 4-sen, (79) and (92) give

$$\underline{\pi}(\text{Poi}, \mathcal{L}(\mathcal{D}), \alpha, \mathcal{D}) \geq \inf_u \left\{ \left( \ln(\alpha) - \frac{1}{2}\ln\left(2u - \frac{1}{2\pi}\int_{-\pi}^{\pi} \ln(2\pi g(\beta))d\beta\right) \right)^+ \right.$$
$$\left. + u - \frac{1}{4\pi}\int_{-\pi}^{\pi} \ln(2\pi e g(\beta))d\beta - \frac{1}{2}\ln\left(2u - \frac{1}{2\pi}\int_{-\pi}^{\pi}\ln(2\pi g(\beta))d\beta\right) \right\} .$$

Posing

$$v = \sqrt{2u - \frac{1}{2\pi}\int_{-\pi}^{\pi} \ln(2\pi g(\beta))d\beta} ,$$

we get that the last infimum is

$$\inf_{v \geq 0} \left\{ (\ln(\alpha) - \ln(v))^+ + \frac{v^2}{2} - \frac{1}{2} - \ln(v) \right\}$$

and we hence get the same function to optimize as in the wgn case. So the exponent is that of formula (65).

We now give a direct analysis of this result. From (20),

$$\begin{aligned} p_e(n) &= \int_{r>0} \left(1 - e^{-\lambda_n V_{\mathcal{E}}^n(r)}\right) g_1^n(r) dr \\ &= \int_0^\infty \left(1 - e^{-\lambda_n V_{\mathcal{E}}^n(v\sqrt{n})}\right) g_1^n(v\sqrt{n}) \sqrt{n} dv . \end{aligned} \qquad (98)$$

In the orthonormal basis which diagonalizes the covariance matrix, we have

$$V_{\mathcal{E}}^n(r) = V_B^n(r) \prod_{i=1}^n \sigma_i.$$

Using the Grenander–Szegö Theorem, we get that

$$\prod_{i=1}^n \sigma_i = e^{n\left[\frac{1}{2\pi} \int_{-\pi}^\pi \ln(g(\beta)) d\beta\right] + o(1)}.$$

Hence if $\mu^n$ is a sequence of Poisson point processes of rates $\lambda_n = e^{nR}$ where

$$R = -h(\mathcal{D}) - \ln(\alpha) = -\frac{1}{2}\ln(2\pi e \alpha^2) - \frac{1}{2}\frac{1}{2\pi}\int_{-\pi}^\pi \ln(g(\beta)) d\beta ,$$

with $\alpha > 1$, then

$$1 - e^{-\lambda_n \prod_{i=1}^n \sigma_i V_B^n(v\sqrt{n})} = e^{-n\left((\ln \alpha - \ln v)^+ + o(1)\right)} .$$

The second term of the integrand in (98) is the same as in the white case (see (64)). So, the exponent $a(v) + b(v)$ of the integrand is indeed the same as in the white Gaussian noise case.

- **Markovian noise.** Under the assumptions of Lemma 1-mn, Theorem 4-sen, (82) and (93) give

$$\underline{\pi}(\text{Poi}, \mathcal{L}(\mathcal{D}), \alpha, \mathcal{D}) \geq \inf_u \left\{ \left(\ln(\alpha) + h(\mathcal{D}) - \psi(u)\right)^+ + u - \phi(u) \right\} , \qquad (99)$$

with

$$\begin{aligned} h(\mathcal{D}) &= h(\mathcal{D}_2|\mathcal{D}_1) \\ \phi(u) &= \sup_{\tau \in \text{SM}_1(\mathbb{R}^2): \ K(\tau||\tau_1 \otimes P) + h(\tau_2|\tau_1) = u} h(\tau_2|\tau_1) \\ \psi(u) &= \sup_{\tau \in \text{SM}_1(\mathbb{R}^2): \ K(\tau||\tau_1 \otimes P) + h(\tau_2|\tau_1) \leq u} h(\tau_2|\tau_1) . \end{aligned}$$

## 8.3 Mismatched Decoding

The setting is that of Section 6.4.

**Theorem 4-sen-Poisson-Mismatch.** *Assume that $\mu^n$ is Poisson with normalized logarithmic intensity $-h(\mathcal{D}) - \ln(\alpha)$ with $\alpha > 1$ and that decoding is MLE. Assume both the design noise $\mathcal{D}$ and the actual noise $\widetilde{\mathcal{D}}$ to be stationary and ergodic and such that the Assumptions H-sen hold. Then the associated error exponent is bounded from below by*

$$\inf_u \left\{ F(u) + \widetilde{I}(u) \right\} , \tag{100}$$

*where $\widetilde{I}(u)$ is the rate function of $\rho_{\widetilde{\mathcal{D}}}^n$ and*

$$F(u) = (\ln(\alpha) + h(\mathcal{D}) - J(u))^+ , \tag{101}$$

*where $J(u) = \sup_{s \leq u}(s - I(s))$ is the volume exponent for $\mathcal{D}$.* □

## 8.4 Matérn Point Process

### 8.4.1 White Noise Case

Throughout this subsection, $\mathcal{D}$ is WN with coordinates with density $f$ on $\mathbb{R}$. As above, $S$ denotes the support of $f$.

Given $x^n \in \mathbb{R}^n$ we associate to it the empirical distribution $\tau_{x^n} \in \mathbb{M}_1(S)$ defined by

$$\tau_{x^n}(A) = \frac{1}{n} \sum_{i=1}^n 1(x_i \in A) .$$

for all Borel sets $A$ of $\mathbb{R}$.

The integrand of the inner integral in the RHS of eqn. (41) depends only on $\tau_{x^n}$. To see this, fix $x^n$ with empirical distribution $\tau = \tau_{x^n}$. We discuss $y^n \in \mathbb{R}^n$ via its conditional empirical distribution given $x^n$, denoted $\gamma = \gamma_{\tau, y^n}$, defined as the regular conditional distribution satisfying,

$$\int_{B \in \mathbb{R}} \int_{A \in S} \gamma(dv \mid u) \tau(du) = \frac{1}{n} \sum_{i=1}^n 1(x_i \in A, y_i \in B) .$$

Then

$$\{\mathbb{D}(y^n, 0) \geq \xi\} \Leftrightarrow \int_{\mathbb{R}^n} \int_S -\ln(f(v)) \gamma(dv \mid u) \tau(du) \geq \xi ,$$

and

$$\{\mathbb{D}(x^n, y^n) \leq \mathbb{D}(x^n, 0)\} \Leftrightarrow \int_{\mathbb{R}^n} \int_S -\ln(f(u-v)) \gamma(dv \mid u) \tau(du) \leq \int_S -\ln(f(u)) \tau(du) .$$

So we can write
$$\int_{y^n \in \mathbb{R}^n} 1_{\mathbb{D}(y^n,0) \geq \xi} 1_{\mathbb{D}(x^n,y^n) \leq \mathbb{D}(x^n,0)} dy^n$$
as $e^{n\psi_n(\tau)}$ with
$$\begin{aligned}\psi_n(\tau) &= \frac{1}{n} \ln \int_{y^n \in \mathbb{R}^n} 1_{\int_{\mathbb{R}^n} \int_S -\ln(f(v))\gamma_{\tau,y^n}(dv|u)\tau(du) \geq \xi} \\ &\quad 1_{\int_{\mathbb{R}^n} \int_S -\ln(f(u-v))\gamma(dv|u)\tau(du) \leq \int_S -\ln(f(u))\tau(du)} dy^n.\end{aligned}$$

From Sanov's theorem, the laws of the random measures $\tau_{x^n}$ satisfy an LDP on the space $(\mathbb{M}_1(S), \mathcal{M}_1(S))$, with the good and convex rate function $K(. \parallel \phi)$ defined in (76).

So, if for all $\nu \in \mathbb{M}_1(S)$,
$$\psi_n(\nu) \to_{n \to \infty} \psi(\nu), \tag{102}$$
then one can expect to deduce from Varadhan's lemma that
$$\underline{\pi}(\text{Mat}, \mathcal{L}(\mathcal{D}), \alpha, \mathcal{D}) \geq \inf_{\tau \in \mathbb{M}_1(S)} \left((-\ln(\alpha) - h(D) - \psi(\tau))^+ + K(\tau \parallel \phi)\right). \tag{103}$$

This is of the same form as our main result in Theorem 4-sen-Poisson.

Notice that the same idea can be applied to all cases where the empirical distribution of the noise satisfies an LDP. For instance, it can be extended to the Markov case by using pairwise-transition empirical distributions.

The main technicality in both cases is the proof of (102). We have no general result on the matter at this stage. We rather look at two special cases below.

### 8.4.2 White Gaussian Noise

Throughout this subsection, $\mathcal{D}$ is WGN with variance $\sigma^2$ and decoding is MLE. Fix $\epsilon > 0$. Consider a sequence of Matérn I processes $\widetilde{\mu}^n$. The point process $\widetilde{\mu}^n$ is built from a Poisson processes $\mu_n$ of rate $\lambda_n = e^{nR}$ where $R = \frac{1}{2} \ln \frac{1}{2\pi e \alpha^2 \sigma^2}$ for $\alpha > 1$, and has for exclusion radius $(\alpha - \epsilon)\sigma\sqrt{n}$. The intensity of this Matérn I point process is
$$\widetilde{\lambda}_n = \lambda_n e^{-\lambda_n V_B^n((\alpha-\epsilon)\sigma\sqrt{n})}$$
and it is easy to see that $\frac{\widetilde{\lambda}_n}{\lambda_n} \to_{n \to \infty} 1$, with $\widetilde{\lambda}_n < \lambda_n$ for all $n$.

Let $\underline{\pi}(\text{Mat}, \mathcal{L}(\text{WGN}), \alpha, \text{WGN})$ denote the error exponent (62) associated with this family of Matérn point processes.

**Theorem 4-wgn-Matérn.** *We have*
$$\underline{\pi}(\text{Mat}, \mathcal{L}(\text{WGN}), \alpha, \text{WGN}) \geq \frac{\alpha^2}{8}, \quad \text{for all } \alpha \geq 2. \tag{104}$$

□

**Proof.** Take an exclusion radius of $(\alpha - \epsilon)\sigma\sqrt{n}$. From Formula (43),

$$p_e(n) \leq \int_{v \in \mathbb{R}^+} \min\left(1, \lambda_n \text{Vol}\left(B^n\left(0, (\alpha - \epsilon)\sigma\sqrt{n}\right)\right)^c \cap B^n\left(y^n(v), (v\sigma\sqrt{n})\right)\right)$$
$$g_1^n(v\sqrt{n})\sqrt{n}dv \ ,$$

with $y^n(v) = (v\sigma\sqrt{n}, 0, \ldots, 0)$. It is shown in Appendix 10.7 that

$$\text{Vol}\left(B^n\left(0, (\alpha - \epsilon)\sigma\sqrt{n}\right)\right)^c \cap B^n\left(y^n(v), (v\sigma\sqrt{n})\right) \leq V_B^n(c(v)\sigma\sqrt{n}) \ , \quad (105)$$

with

$$c(v) = \begin{cases} 0 & \text{if } 0 < v < \frac{\widetilde{\alpha}}{2} \\ \sqrt{v^2 - (v - \frac{\widetilde{\alpha}^2}{2v})^2} & \text{if } \frac{\widetilde{\alpha}}{2} < v < \frac{\widetilde{\alpha}}{\sqrt{2}} \\ v & \text{if } \frac{\widetilde{\alpha}}{\sqrt{2}} < v \ . \end{cases} \quad (106)$$

with $\widetilde{\alpha} = \alpha - \epsilon$. Hence, by the same arguments as in Theorem 4-wgn-Poisson,

$$\underline{\pi}(\text{Mat}, \mathcal{L}(\text{WGN}), \alpha, \text{WGN}) \geq \inf_{v > 0} b(v) + a(v) \ ,$$

with $a(v) = \frac{v^2}{2} - \frac{1}{2} - \ln v$ and

$$b(v) = \begin{cases} \infty & \text{if } 0 < v < \frac{\widetilde{\alpha}}{2} \\ \ln \alpha - \frac{1}{2}\ln(v^2 - (v - \frac{\widetilde{\alpha}^2}{2v})^2) & \text{if } \frac{\widetilde{\alpha}}{2} < v < \frac{\widetilde{\alpha}}{\sqrt{2}} \\ (\ln \alpha - \ln v)^+ & \text{if } \frac{\widetilde{\alpha}}{\sqrt{2}} < v \ . \end{cases} \quad (107)$$

The announced result follows when minimizing over $v$ for each $\widetilde{\alpha} \geq 2$ and then letting $\epsilon$ to 0. $\square$

**Remark 9.** The lower bound on $\underline{\eta}(\alpha)$ given in the two last theorems, namely

$$\pi(\alpha) = \begin{cases} \frac{\alpha^2}{2} - \frac{1}{2} - \ln \alpha & \text{if } 1 \leq \alpha < \sqrt{2} \\ \frac{1}{2} - \ln 2 + \ln \alpha & \text{if } \sqrt{2} \leq \alpha < 2 \\ \frac{\alpha^2}{8} & \text{if } \alpha \geq 2 \ . \end{cases} \quad (108)$$

was first obtained by Poltyrev [17] (see eqns. (32) and (36) therein) and is referred to as *Poltyrev's error exponent*. The Poltyrev random coding error exponent is defined to be the part of this function which is comprised of the two portions $1 \leq \alpha < \sqrt{2}$ and $\sqrt{2} \leq \alpha < 2$. So the Poisson point process achieves this Poltyrev random coding error exponent in this range, whereas the Matérn I point process achieves the Poltyrev function in the range $\alpha \geq 2$.

### 8.4.3 Symmetric Exponential Noise

This section is focused on the case where $\mathcal{D}$ is white symmetric exponential noise and on the Matérn-$\mathcal{D}$-$\xi$ point process, which is a hard exclusion process where the exclusion regions are $L_1$ balls of radius

$$r^n(\xi) = \frac{n\sigma}{\sqrt{2}}(\xi - \ln(\sqrt{2}\sigma)) \ ,$$

for $\xi > \ln(\sqrt{2}\sigma)$. In what follows, we build this Matérn point process, $\widetilde{\mu}^n$, from a Poisson point process $\mu^n$ of intensity $\lambda_n = e^{nR}$ with $R = -\ln(\sqrt{2}e\sigma\alpha)$, where $\alpha > 1$. The parameter $\xi$ is chosen as follows:

$$\xi = \alpha - \epsilon + \ln(\sqrt{2}\sigma),$$

so that the $L_1$ exclusion radius is

$$r_n = \frac{n\sigma}{\sqrt{2}}(\alpha - \epsilon).$$

The intensity of the associated Matérn point process is then

$$\widetilde{\lambda}_n = \lambda_n e^{-\lambda_n V_{B,1}^n(r_n)},$$

with

$$V_{B,1}^n(r_n) = \frac{(2r_n)^n}{n!} = \frac{(\sqrt{2}\sigma(\alpha - \epsilon))^n n^n}{n!}$$

the volume of the $L_1$ ball of radius $r_n$. It is easy to see that $\widetilde{\lambda}_n \leq \lambda_n$ for all $n$ and that $\lim_{n\to\infty} \frac{\widetilde{\lambda}_n}{\lambda_n} = 1$.

By following the same line of thought as in the proof of Corollary 3-wgn-Matérn, we deduce from (41) that

$$p_e(n) \leq \int_{r>0} \min\left(1, \lambda_n \sup_{x^n:|x^n|_1=r} \text{Vol}\left(B_1^n(0, r_n)^c \cap B_1^n(x^n, r)\right)\right) g_\sigma^n(r)dr, \quad (109)$$

where $|.|_1$ denotes the $L_1$ norm, $B_1^n(x, r)$ the $L_1$ ball of center $x$ and radius $r$ and $g_\sigma^n(r)$ here denotes the density of the $L_1$ norm of $D^n$, which is $\Gamma(\theta, n)$ with $\theta = \frac{\sqrt{2}}{\sigma}$, that is

$$g_\sigma^n(r) = e^{-\frac{\sqrt{2}}{\sigma}r}\left(\frac{\sqrt{2}}{\sigma}\right)^n \frac{r^{n-1}}{\Gamma(n)}, \quad r \geq 0.$$

So, when posing $r = vn\sigma/\sqrt{2}$, we get that the R.H.S. of (109) is

$$\int_{r>0} \min\left(1, \lambda_n \sup_{x^n:|x^n|_1=\frac{vn\sigma}{\sqrt{2}}} W(x^n, v)\right) e^{-vn} \frac{(vn)^n}{v\Gamma(n)}dv, \quad (110)$$

with

$$W(x^n, v) = \text{Vol}\left(B_1^n\left(0, \frac{n\sigma(\alpha-\epsilon)}{\sqrt{2}}\right)^c \cap B_1^n\left(x^n, \frac{vn\sigma}{\sqrt{2}}\right)\right). \quad (111)$$

Let $\widetilde{\alpha} = \alpha - \epsilon$. If $v \leq \frac{\widetilde{\alpha}}{2}$, then $W(x^n, v) = 0$ for all $x^n$ with $|x^n|_1 = \frac{vn\sigma}{\sqrt{2}}$. It is proved in Appendix 10.8 that if $v > \frac{\widetilde{\alpha}}{2}$, then

$$\lim_{n\to\infty} \frac{1}{n} \ln\left(\sup_{x^n:|x^n|_1=\frac{vn\sigma}{\sqrt{2}}} W(x^n, v)\right) = \ln(\sqrt{2}ve\sigma). \quad (112)$$

So, by the same arguments as in the WGN case, the error exponent associated with this sequence of Matérn point processes satisfies the bound:

$$\underline{\pi}(\text{Mat}, \mathcal{L}(\mathcal{D}), \alpha, \mathcal{D}) \geq \inf_{v>0} b(v) + a(v) ,$$

with $a(v) = v - \ln(v) - 1$, (stemming from $e^{-vn} \frac{(vn)^n}{v\Gamma(n)}$), and

$$b(v) = \begin{cases} \infty & \text{if } 0 < v < \frac{\tilde{\alpha}}{2} \\ (\ln \tilde{\alpha} - \ln v)^+ & \text{if } \frac{\tilde{\alpha}}{2} < v . \end{cases}$$

(stemming from $\min\left(1, \lambda_n \sup_{x^n:|x^n|_1=\frac{v\sigma n}{\sqrt{2}}} W(x^n, v)\right)$ in (110)). This leads to:

**Theorem 4-symmetric-exponential-Matérn.** *For $\mathcal{D}$ symmetric exponential, we have*

$$\underline{\pi}(\text{Mat}, \mathcal{L}(\mathcal{D}), \alpha, \mathcal{D}) \geq \begin{cases} \alpha - \ln(\alpha) - 1 & \text{for } \alpha \leq 2 \\ \ln(\alpha) + 1 - 2\ln(2) & \text{for } 2 \leq \alpha \leq 4 \\ \frac{\alpha}{2} - \ln(\alpha) - 1 + 2\ln(2) & \text{for } \alpha \geq 4. \end{cases} \quad (113)$$

□

## 9 The Channel with Power Constraints

### 9.1 The AWGN Channel

Consider communication over the AWGN channel with noise variance $\sigma^2$ per degree of freedom using codewords of block length $n$ subject to the power constraint $P$.

We write $A^2$ for the *signal-to-noise ratio* $P/\sigma^2$. A *code* is a finite subset $\mathcal{T}$ of points in $B^n(0, \sqrt{nP})$ [2]. The elements of the code are called *codewords*. We call $R(\mathcal{T}) = \frac{1}{n} \ln |\mathcal{T}|$ the *rate* of the code. Consider the Voronoi decomposition of $\mathbb{R}^n$ determined by the codewords. For each codeword, the probability that a Gaussian random vector in $\mathbb{R}^n$, centered at the codeword with independent coordinates each having variance $\sigma^2$, lands outside the corresponding Voronoi cell, is called the associated probability of error. By the average probability of error of the code, denoted $p_e(\mathcal{T})$, we mean the average over the codewords of the corresponding probability of error. Given $\mathcal{T}$, this only depends on the signal-to-noise ratio $A^2$. One of the central problems of coding for the AWGN channel is that of understanding, for each fixed $A > 0$, the set of achievable pairs $(R(\mathcal{T}), p_e(\mathcal{T}))$, as one varies $\mathcal{T}$.

---

[2]Another option consists in taking the codewords in $SS^{n-1}(\sqrt{nP})$. The two options lead to the same results.

### 9.1.1 Shannon Capacity

For fixed $A > 0$, given $R > 0$, we may ask whether there exists a sequence of codes $\mathcal{T}_n$ in $\mathbb{R}^n$, $n \geq 1$, having rate at least $R$, asymptotically as $n \to \infty$, and such that $p_e(\mathcal{T}_n) \to 0$ as $n \to \infty$. As proved by Shannon, this is possible if $R < \frac{1}{2}\ln(1 + A^2)$ and not possible if $R > \frac{1}{2}\ln(1 + A^2)$. Thus $\frac{1}{2}\ln(1 + A^2)$ may be called the Shannon capacity of the AWGN channel.

### 9.1.2 The Error Exponent

Let $A > 0$ and $0 < R \leq \frac{1}{2}\ln(1 + A^2)$. Let $P_{e,opt}(n, R, A)$ denote the infimum of $p_e(\mathcal{T})$ over all codes in $\mathbb{R}^n$ of rate at least $R$ when the signal-to-noise ratio is $A^2$. We write

$$\mathcal{E}_{\text{exact}}(n, R, A) = -\frac{1}{n}\log P_{e,opt}(n, R, A) \ .$$

$$\text{Let} \quad \bar{\mathcal{E}}_{\text{exact}}(R, A) = \limsup_n \mathcal{E}_{\text{exact}}(n, R, A), \text{ and}$$
$$\underline{\mathcal{E}}_{\text{exact}}(R, A) = \liminf_n \mathcal{E}_{\text{exact}}(n, R, A). \tag{114}$$

Assuming these are identical, we denote this common limit by $\mathcal{E}_{\text{exact}}(R, A)$. For fixed $A > 0$ the function $R \mapsto \mathcal{E}_{\text{exact}}(R, A)$ is called the *error exponent* or the *reliability function* of the AWGN channel when the signal-to-noise ratio is $A^2$.

The best currently known upper and lower bounds for the error exponent are given in [2], to which we refer the reader.[3] We write $R \mapsto \mathcal{E}_{SG}(R, A)$ for the best known lower bound, since it is due to Shannon [19] and Gallager [8]. We write $R \mapsto \mathcal{E}_{ABL}(R, A)$ for the best known upper bound, since it is due to Ashikhmin, Barg and Litsyn [2]. These bounds coincide, and thus $\mathcal{E}_{\text{exact}}(R, A)$ is known, for

$$\frac{1}{2}\ln\left(\frac{1}{2} + \frac{A}{4} + \frac{1}{2}\sqrt{1 + \frac{A^2}{4}}\right) \leq R \leq \frac{1}{2}\ln(1 + A^2) \ ,$$

but there is a gap between these bounds for all lower rates.

## 9.2 The Additive Stationary Ergodic Noise Channel

The setting is the same as above but for the noise vector which is not i.i.d. Gaussian anymore. Here, the noise vector for dimension $n$, $D^n = (\mathcal{D}_1, \ldots, \mathcal{D}_n)$, has its coordinates which are the $n$ first values of a discrete time, stationary and ergodic, real–valued stochastic process $\{\mathcal{D}_k\}_{k \geq 1}$. We assume that $\mathcal{D}_0$ is centered and has a finite variance. We also assume that $D^n$ has a density on $\mathbb{R}^n$ for all $n$ denoted by $f^n$.

---

[3] Note that the upper bounds are on $\bar{\mathcal{E}}_{\text{exact}}(R, A)$ and the lower bounds are on $\underline{\mathcal{E}}_{\text{exact}}(R, A)$.

### 9.2.1 The Shannon Capacity

We assume here that the codewords belong to $B^n(0, \sqrt{nP})$. The Shannon capacity of the additive channel with this noise and under this restriction is known to be

$$C_P(\mathcal{D}) = \lim_{n \to \infty} \frac{1}{n} \sup_{T^n,\ \mathbb{E}(\sum_{i=1}^n (T_i^n)^2) < nP} I(T^n, T^n + D^n), \qquad (115)$$

where the supremum bears on all distribution functions for $T^n = (T_1^n, \ldots, T_n^n) \in \mathbb{R}^n$ such that $\mathbb{E}(\sum_{i=1}^n (T_i^n)^2) < nP$.

### 9.2.2 Connections between Capacities with and without Restrictions

Let $\sigma^2$ denote the variance of $\mathcal{D}(0)$. The relation between the Shannon capacity and that defined in Theorems 1-sen and 2-sen is given by the following lemma (which is due to Shannon – we give a proof for the sake of self containedness).

**Lemma 5.** *Under the foregoing assumptions,*

$$\frac{1}{2} \ln(2\pi eP) + c(\mathcal{D}) \leq C_P(\mathcal{D}) \leq \frac{1}{2} \ln(2\pi e(P + \sigma^2)) + c(\mathcal{D}) \qquad (116)$$

*and*

$$C_P(\mathcal{D}) = \frac{1}{2} \ln(2\pi eP) + c(\mathcal{D}) + O(1/P), \qquad (117)$$

*when $P$ tends to infinity.* □

**Proof.** We have

$$I(T^n + \mathcal{D}^n; T^n) = h(T^n + \mathcal{D}^n) - h(T^n + \mathcal{D}^n \mid T^n) = h(T^n + \mathcal{D}^n) - h(\mathcal{D}^n).$$

It is well known that for all stationary sequences $A_1, A_2, \ldots$

$$h(A_1, A_2, \ldots, A_n) \leq \frac{n}{2} \ln(2\pi e \mathrm{Var}(A_1)).$$

Hence

$$\frac{1}{n} I(T^n + \mathcal{D}^n; T^n) \leq \frac{1}{2} \ln(2\pi e(P + \sigma^2)) - \frac{1}{n} h(\mathcal{D}^n).$$

This gives the upper bound and the fact that

$$C_P(\mathcal{D}) \leq \frac{1}{2} \ln(2\pi eP) - h(\mathcal{D}) + O(1/P),$$

when $P$ tends to infty.

For the lower bound, we use the inequality

$$h(T^n + \mathcal{D}^n) \geq h(T^n)$$

to deduce that

$$I(T^n + \mathcal{D}^n; T^n) = h(T^n + \mathcal{D}^n) - h(T^n + \mathcal{D}^n \mid T^n) \geq h(T^n) - h(\mathcal{D}^n).$$

Taking now $T^n$ Gaussian with i.i.d. $\mathcal{N}(0, P)$ coordinates, we get that

$$C_P(\mathcal{D}) \geq \frac{1}{2} \ln(2\pi e P) - h(\mathcal{D}).$$

$\square$

### 9.2.3 Error Exponents with Restrictions

In the case with general noise, one defines

$$\mathcal{E}(n, R, P, \mathcal{D}) = -\frac{1}{n} \log p_{e,opt}(n, R, P, \mathcal{D}), \qquad (118)$$

with $p_{e,opt}(n, R, P, \mathcal{D})$ the infimum of $p_e(\mathcal{T})$ over all codes in $\mathbb{R}^n$ of rate at least $R$ and all decoding rules, when the signal power is $P$ and the noise is $\mathcal{D}$. One then defines

$$\bar{\mathcal{E}}(R, P, \mathcal{D}) = \limsup_n \mathcal{E}(n, R, P, \mathcal{D}), \text{ and}$$
$$\underline{\mathcal{E}}(R, P, \mathcal{D}) = \liminf_n \mathcal{E}(n, R, P, \mathcal{D}).$$

Assuming these are identical, one denotes this common limit by $\mathcal{E}(R, P, \mathcal{D})$.

Here is the relation between the notation of the Gaussian case:

$$\mathcal{E}_{\text{exact}}(R, A) = \mathcal{E}(R, P, \text{WGN}(\sigma^2))$$

with $A = P/\sigma^2$.

### 9.2.4 Connections between Error Exponents with and without Restrictions

**Theorem 5-sen.** *Let $\mathcal{D}$ be a stationary and ergodic noise process and let $\alpha > 1$. For all families of stationary and ergodic point processes $\mu = \{\mu^n\}$, each with normalized logarithmic density bounded from above by $-h(\mathcal{D}) - \ln(\alpha)$, for all families of jointly stationary decoding regions $\mathcal{C} = \{\mathcal{C}^n\}$ and for all $P > 0$ such that $\frac{1}{2}\ln(2\pi e P) > h(\mathcal{D}) + \ln(\alpha)$, we have*

$$\underline{\mathcal{E}}\left(\frac{1}{2}\ln(2\pi e P) - h(\mathcal{D}) - \ln(\alpha), P, \mathcal{D}\right) \geq \underline{\pi}(\mu, \mathcal{C}, \alpha-, \mathcal{D}), \qquad (119)$$

*and*

$$\underline{\mathcal{E}}\left(C_P(\mathcal{D}) - \ln(\alpha) - \frac{1}{2}\ln\left(1 + \frac{\sigma^2}{P}\right), P, \mathcal{D}\right) \geq \underline{\pi}(\mu, \mathcal{C}, \alpha-, \mathcal{D}), \qquad (120)$$

*where*

$$\underline{\pi}(\mu, \mathcal{C}, \alpha-, \mathcal{D}) = \lim_{\epsilon \downarrow 0} \underline{\pi}(\mu, \mathcal{C}, \alpha - \epsilon, \mathcal{D}).$$

with $\underline{\pi}(\mu, \mathcal{C}, x, \mathcal{D})$ *the error exponent without restriction defined in (62). In addition*

$$\liminf_{P \to \infty} \underline{\mathcal{E}}(C_P(\mathcal{D}) - \ln(\alpha), P, \mathcal{D}) \geq \underline{\pi}(\mu, \mathcal{C}, \alpha-, \mathcal{D}). \tag{121}$$

□

**Proof.** From the very definition of Palm probabilities, for all $n$,

$$p_e^{pp}(n, \mu^n, \mathcal{C}^n, \alpha, \mathcal{D}) = \frac{\mathbb{E}^n \left( \sum_{k \text{ s.t. } T_k^n \in B^n(0, \sqrt{nP})} p_{e,k} \right)}{e^{-nh(\mathcal{D})} e^{-n \ln(\alpha)} V_B^n(\sqrt{nP})},$$

where $p_{e,k}$ denotes the probability that $T_k^n + D_k^n$ does not belong to $\mathcal{C}_k^n$ given $\{T_l^n, \mathcal{C}_l^n\}_l$. Hence, for all $\gamma > 0$,

$p_e^{pp}(n, \mu^n, \mathcal{C}^n, \alpha, \mathcal{D})$

$$\geq \frac{\mathbb{E}^n \sum_{k \text{ s.t. } T_k^n \in B^n(0, \sqrt{nP})} p_{e,k} 1_{\mu^n(B^n(0, \sqrt{nP})) \geq (2\pi eP)^{\frac{n}{2}} e^{-nh(\mathcal{D})} e^{-n \ln(\alpha+\gamma)}}}{e^{-nh(\mathcal{D})} e^{-n \ln(\alpha)} V_B^n(\sqrt{nP})}$$

$$\geq \mathbb{P}^n \left( \mu^n(B^n(0, \sqrt{nP})) \geq (2\pi eP)^{\frac{n}{2}} e^{-nh(\mathcal{D})} e^{-n \ln(\alpha+\gamma)} \right)$$

$$p_{e,opt}(n, \frac{1}{2} \ln(2\pi eP) - h(\mathcal{D}) - \ln(\alpha + \gamma), P, \mathcal{D}) e^{-n \ln(\alpha+\gamma)} e^{n \ln(\alpha)} \frac{(2\pi eP)^{\frac{n}{2}}}{V_B^n(\sqrt{nP})}.$$

Hence

$$-\frac{1}{n} \ln \left( p_e^{pp}(n, \mu^n, \mathcal{C}^n, \alpha, \mathcal{D}) \right)$$

$$\leq -\frac{1}{n} \ln \left( p_{e,opt}(n, \frac{1}{2} \ln(2\pi eP) - h(\mathcal{D}) - \ln(\alpha + \gamma), P, \mathcal{D}) \right)$$

$$-\frac{1}{n} \ln \left( \mathbb{P}^n \left( \mu^n(B^n(0, \sqrt{nP})) \geq (2\pi eP)^{\frac{n}{2}} e^{-nh(\mathcal{D})} e^{-n(\alpha+\gamma)} \right) \right)$$

$$- \ln(\alpha) + \ln(\alpha + \gamma) - \frac{1}{n} \ln \left( \frac{(2\pi eP)^{\frac{n}{2}}}{V_B^n(\sqrt{nP})} \right).$$

When taking a liminf in $n$, we get that

$$\underline{\pi}(\mu, \mathcal{C}, \alpha, \mathcal{D}) \leq \ln(1 + \gamma/\alpha)) + \underline{\mathcal{E}} \left( \frac{1}{2} \ln(2\pi eP) - h(\mathcal{D}) - \ln(\alpha + \gamma), P, \mathcal{D} \right). \tag{122}$$

Equivalently, for all $\tilde{\alpha} > 1$ for all $\tilde{\gamma}$ small enough,

$$\underline{\pi}(\mu, \mathcal{C}, \tilde{\alpha}-\tilde{\gamma}, \mathcal{D}) \leq \ln(\tilde{\alpha})/(\tilde{\alpha}-\tilde{\gamma})) + \underline{\mathcal{E}}\left(\frac{1}{2}\ln(2\pi e P) - h(\mathcal{D}) - \ln(\tilde{\alpha}), P, \mathcal{D}\right) \quad (123)$$

and (119) follows when letting $\tilde{\gamma}$ to 0 in the last relation.

We get (120) from (119) when using the second inequality of (116) and the fact that the function $x \to \underline{\mathcal{E}}(x, P, \mathcal{D})$ is non-increasing.

By the same monotonicity arguments, we get from (122) that for $P$ large enough and for $\gamma$ small enough,

$$\underline{\pi}(\mu, \mathcal{C}, \alpha, \mathcal{D}) \leq \ln(1+\gamma/\alpha)) + \underline{\mathcal{E}}\left(C_P(\mathcal{D}) - \ln(\alpha+\gamma) - \frac{1}{2}\ln\left(1+\frac{\sigma^2}{P}\right), P, \mathcal{D}\right) \ .$$

Hence, for all $\tilde{\alpha} > 1$, for all $P$ large enough and $\tilde{\gamma}$ small enough,

$$\underline{\pi}\left(\mu, \mathcal{C}, \frac{\tilde{\alpha}-\tilde{\gamma}}{\sqrt{1+\frac{\sigma^2}{P}}}, \mathcal{D}\right) \leq \ln(\tilde{\alpha}/(\tilde{\alpha}-\tilde{\gamma})) + \underline{\mathcal{E}}(C_P(\mathcal{D}) - \ln(\tilde{\alpha}), P, \mathcal{D})$$

and (121) follows by letting $\tilde{\gamma}$ tend to 0 and $P$ tend to $\infty$. $\square$

As a direct corollary of the last theorem, we have

$$\underline{\mathcal{E}}\left(C_P(\mathcal{D}) - \ln(\alpha) - \frac{1}{2}\ln\left(1+\frac{\sigma^2}{P}\right), P, \mathcal{D}\right) \geq \underline{\pi}(\text{Poi}, \mathcal{L}(\mathcal{D}), \alpha-, \mathcal{D}) \quad (124)$$

$$\underline{\mathcal{E}}\left(C_P(\mathcal{D}) - \ln(\alpha) - \frac{1}{2}\ln\left(1+\frac{\sigma^2}{P}\right), P, \mathcal{D}\right) \geq \underline{\pi}(\text{Mat}, \mathcal{L}(\mathcal{D}), \alpha-, \mathcal{D}) \quad (125)$$

with $\mathcal{L}(\mathcal{D})$ the MLE decomposition for $\mathcal{D}$.

### 9.2.5 AWGN Channel

Given $A = P/\sigma^2 > 0$ and $\alpha \geq 1$, consider $\underline{\mathcal{E}}_{\text{exact}}(\frac{1}{2}\ln\frac{1+A^2}{\alpha^2}, A)$, where the function $R \mapsto \underline{\mathcal{E}}(R, A)$ is defined in subsection 9.1.2. Then, it follows from Theorem 5-sen that, for all $\alpha \geq 1$,

$$\lim_{A \to \infty} \underline{\mathcal{E}}_{\text{exact}}\left(\frac{1}{2}\ln\frac{1+A^2}{\alpha^2}, A\right) \geq \underline{\eta}(\alpha, \sigma^2), \quad (126)$$

with $\underline{\eta}(\alpha, \sigma^2)$ the error exponent defined in Section 9.1.2. Hence for all $\alpha \geq 1$,

$$\lim_{A \to \infty} \underline{\mathcal{E}}_{\text{exact}}\left(\frac{1}{2}\ln\frac{1+A^2}{\alpha^2}, A\right) \geq \pi(\alpha) \ , \quad (127)$$

with $\pi(\alpha)$ the Poltyrev exponent (108).

# 10 Appendix

## 10.1 Density of the Norm of the Gaussian Vector

We recall that the $\Gamma(\theta, \nu)$ density is of the form
$$1_{x>0} e^{-\theta x} \theta^\nu x^{\nu-1} \frac{1}{\Gamma(\nu)}.$$

If $\sigma = 1$, the density of $D_1^2 + \cdots + D_n^2$ is $\Gamma(1/2, n/2)$, that is
$$f(x) = 1_{x>0} e^{-x/2} \frac{1}{2^{n/2}} x^{n/2-1} \frac{1}{\Gamma(n/2)}.$$

Hence by the change of variable formula for densities, the density of $\sqrt{D_1^2 + \cdots + D_n^2}$ at $r$ is
$$g_1^n(r) = 1_{r>0} e^{-r^2/2} \frac{1}{2^{n/2}} r^{n-1} \frac{2}{\Gamma(n/2)}.$$

## 10.2 Symmetric Exponential Distribution

The differential entropy of the symmetric exponential distribution of variance $\sigma^2$ is $h(D) = \ln(\sqrt{2}e\sigma)$.

We prove below that among the distributions on $\mathbb{R}$ such that the mean of the absolute value is fixed, the distribution with maximal entropy is symmetric exponential.

The differential entropy $h$ of a real–valued random variable with density $\tau$ and with mean absolute value $t$ is $\tau^+ h^+ + \tau^- h^- + h(\tau^+)$ with $h(x) = x\ln(x) + (1-x)\ln(1-x)$,
$$\tau^+ = \int_0^\infty \tau(x) dx, \quad \tau^- = \int_{-\infty}^0 \tau(x) dx$$

and
$$h^+ = -\int_0^\infty \ln(\tau(x)/\tau^+) \tau(x)/\tau^+ dx, \quad h^- = -\int_{-\infty}^0 \ln(\tau(x)/\tau^-) \tau(x)/\tau^- dx.$$

Let
$$t^+ = \int_0^\infty \tau(x)/\tau^+ x dx, \quad t^- = -\int_{-\infty}^0 \tau(x)/\tau^- x dx.$$

We have $\tau^+ t^+ + \tau^- t^- = t$. The maximum of entropy distribution among the set of distributions on $\mathbb{R}^+$ with mean $a$ is the exponential of parameter $a$. Hence, for $t^+$ fixed, the maximum entropy is reached for $\tau(x)/\tau^+$ exponential of mean $t^+$. Using a similar argument for $\tau(x)/\tau^-$, we get that the maximal entropy is obtained for the parameters $\tau^+ = v^*$ and $t^+ = u^*$ given by:

$$(u^*, v^*) = \mathrm{argmax}_{u \in [0,t], v \in [0,1]} v \ln(eu) + (1-v)\ln(e(t-uv)/(1-v)) + h(v).$$

For fixed $v$, the optimal $u$ is $u^* = t$. At this value the function to be optimized in $v$ is $\ln(et) + h(v)$ which is maximal in $v^* = 1/2$.

## 10.3 Direct Proof of Theorem 4-wgn-Poisson

Theorem 4-wgn-Poisson is a special case of Theorem 4-sen-Poisson (see the WGN example following the proof of Theorem 4-sen-Poisson). We nevertheless give a direct proof below.

We first get an upper bound on

$$g_1^n(v\sqrt{n}) = \mathbb{1}_{v>0} e^{-nv^2/2} v^{n-1} \frac{n^{n/2}}{\sqrt{n}} \frac{1}{2^{n/2}} \frac{2}{\Gamma(n/2)}.$$

Using the bound

$$K\left(\frac{x}{e}\right)^x \sqrt{\frac{2\pi}{x}} \geq \Gamma(x) \geq \left(\frac{x}{e}\right)^x \sqrt{\frac{2\pi}{x}}, \tag{128}$$

for some constant $K > 1$, we get

$$\frac{n^{n/2}}{\sqrt{n}} \frac{1}{2^{n/2}} \frac{2}{\Gamma(n/2)} \leq \frac{n^{n/2}}{\sqrt{n}} \frac{1}{2^{n/2}} \frac{\sqrt{n}}{\sqrt{\pi}(n/2e)^{n/2}} = \frac{1}{\sqrt{\pi}} \frac{1}{e^{-n/2}}.$$

Hence

$$\sqrt{n} g_1^n(v\sqrt{n}) \leq \frac{\sqrt{n}}{\sqrt{\pi}} \frac{1}{v} e^{-n\left(\frac{v^2}{2} - \frac{1}{2} - \ln(v)\right)}. \tag{129}$$

The bounds (128) can also be used to show that there exist constants $C$ and $C'$ such that for all $n$ large enough for all $v > 0$,

$$(2e\pi v)^{\frac{n}{2}} \frac{C}{\sqrt{2+n}} \leq V_B^n(\sqrt{n}v) \leq (2e\pi v)^{\frac{n}{2}} \frac{C'}{\sqrt{2+n}} \tag{130}$$

We now get an upper bound on $1 - e^{-\lambda_n V_B^n(v\sigma\sqrt{n})}$. For $x \geq 0$, $1 - e^{-x} \leq \min(x, 1)$. Hence

$$1 - e^{-\lambda_n V_B^n(v\sigma\sqrt{n})} \leq \min(\lambda_n V_B^n(v\sigma\sqrt{n}), 1)$$

and for $n$ large enough,

$$\frac{C}{\sqrt{n+2}} \left(\frac{v}{\alpha}\right)^n \leq \lambda_n V_B^n(v\sigma\sqrt{n}) \leq \frac{C'}{\sqrt{n+2}} \left(\frac{v}{\alpha}\right)^n. \tag{131}$$

Let $a_n$ denote the real number such that

$$\lambda_n V_B^n(a_n \sigma \sqrt{n}) = 1.$$

For $n$ large enough, $a_n > \alpha$. In addition $a_n \to \alpha$ as $n \to \infty$. Thus, from (64) and (131),

$$p_e(n) \leq C' \int_0^{a_n} e^{-n\left(\frac{v^2}{2} - \frac{1}{2} - 2\ln(v) + \ln(\alpha)\right)} \frac{1}{v} dv$$
$$+ \sqrt{\frac{n}{\pi}} \int_{a_n}^{\infty} e^{-n\left(\frac{v^2}{2} - \frac{1}{2} - \ln(v)\right)} \frac{1}{v} dv. \tag{132}$$

The minimum of the function

$$h_1(v) = \frac{v^2}{2} - \frac{1}{2} - 2\ln(v) + \ln(\alpha)$$

is reached for $v = \sqrt{2}$ whereas that of the function

$$h_2(v) = \frac{v^2}{2} - \frac{1}{2} - \ln(v)$$

is reached for $v = 1$.

If $\alpha \geq \sqrt{2}$, the minimum of the function $h_1(v)$ is reached over the $[0, \alpha]$ interval and the dominant term of the first integral is

$$e^{-nh_1(\sqrt{2})} = e^{-n(\frac{1}{2} - \ln(2) + \ln(\alpha))}.$$

Since the function $h_1(v)$ is increasing on $(1, \infty]$, the dominant term in the second one is less than

$$e^{-nh_2(\sqrt{2})} = e^{-n(\frac{1}{2} - \ln(2))}.$$

From (132), defining $\hat{h}_1(v) = h_1(v) - h_1(\sqrt{2})$ and $\hat{h}_2(v) = h_2(v) - h_1(\sqrt{2})$, we get

$$-\frac{1}{n}\ln(p_e(n)) \geq \left(\frac{1}{2} - \ln(2) + \ln(\alpha)\right) + \beta(n)$$

with

$$\beta(n) = -\frac{1}{n}\ln\left(\frac{e}{\pi}\int_0^{a_n}\frac{e^{-n\hat{h}_1(v)}}{v}dv + \frac{\sqrt{n}}{\pi}\int_{a_n}^{\infty}\frac{e^{-n\hat{h}_2(v)}}{v}dv\right).$$

We now conclude the proof by showing that each of the positive functions

$$I_1(n) = C'\int_0^{a_n}\frac{e^{-n\hat{h}_1(v)}}{v}dv, \quad I_2(n) = \frac{\sqrt{n}}{\pi}\int_{a_n}^{\infty}\frac{e^{-n\hat{h}_2(v)}}{v}dv$$

is bounded from above by a constant when $n$ tends to infinity. By dominated convergence, $I_2(n) \to_{n\to\infty} 0$. For all $0 < c < \sqrt{2}$, we have

$$I_1(n) \leq C'\frac{2c}{\sqrt{2} - c} + \int_{[0,a_n]\setminus[\sqrt{2}-c,\sqrt{2}+c]}\frac{e^{-n\hat{h}_1(v)}}{v}dv,$$

where the last integral tends to 0 by dominated convergence. This concludes the proof in this case.

For $\alpha < \sqrt{2}$, the dominant term of the first integral is

$$e^{-nh_1(a_n)} = e^{-n(\frac{1}{2}a_n^2 - \frac{1}{2} - 2\ln(a_n) + \ln(\alpha))}.$$

Since the function $h_2(v)$ is increasing for $v \geq 1$, the dominant term of the second one is

$$e^{-nh_2(a_n)} = e^{-n(\frac{1}{2}a_n^2 - \frac{1}{2} - \ln(a_n))}.$$

We then proceed as above. From (132), defining $\hat{h}_1(v) = h_1(v) - h_1(\alpha)$ and $\hat{h}_2(v) = h_2(v) - h_2(\alpha)$, with

$$h_1(\alpha) = h_2(\alpha) = \frac{\alpha^2}{2} - \frac{1}{2} - \ln(\alpha),$$

we get

$$-\frac{1}{n}\ln(p_e(n)) \geq \left(\frac{\alpha^2}{2} - \frac{1}{2} - \ln(\alpha)\right) + \beta(n)$$

with $\beta(n)$ as above (with these new functions) and it is enough to show that each of the positive functions

$$I_1(n) = C' \int_0^{a_n} \frac{e^{-n\hat{h}_1(v)}}{v} dv, \quad I_2(n) = \frac{\sqrt{n}}{\pi} \int_{a_n}^{\infty} \frac{e^{-n\hat{h}_2(v)}}{v} dv$$

is bounded from above by a function $g(n)$ such that $\frac{1}{n}\ln(g(n))$ tends to 0 when $n$ tends to infinity to complete the proof. For all $0 < c < \alpha$, and for $n$ large enough,

$$I_1(n) \leq C' \frac{2c}{\alpha - c} + \int_{[0,a_n]\setminus[\alpha-c,\alpha+c]} \frac{e^{-n\hat{h}_1(v)}}{v} dv,$$

where the last integral tends to 0 by dominated convergence. Similarly, for $n$ large enough,

$$I_2(n) \leq \frac{\sqrt{n}}{\pi} \frac{2c}{\alpha - c} + \int_{[a_n,\infty]\setminus[\alpha-c,\alpha+c]} \frac{e^{-n\hat{h}_2(v)}}{v} dv,$$

where the last integral tends to 0 by dominated convergence. Hence one can take $g(n) = K\sqrt{n}$ and this concludes the proof. $\square$

## 10.4 Completion of the Proof of Lemma 1-mn

We have to prove that if $\{A_n\}$ and $\{B_n\}$ are real-valued sequences that both satisfy an LDP with the same good and convex rate function $I$, then all sequences $\{X_n\}$ such that

$$A_n \leq X_n \leq B_n, \quad \forall n$$

satisfy an LDP with the rate function $I$ under the assumptions of the lemma.

First apply Varadhan's integral lemma to conclude that for every real $\theta$ we have

$$\lim_n \frac{1}{n} \ln E[e^{\theta A_n}] = \Lambda(\theta) = \sup_x(\theta x - I(x))$$

and the same for $B_n$. Notice that $\Lambda(\theta)$ is necessarily lower-semicontinuous (since it is a Fenchel-Legendre transform).

From this we see (by separately considering $\theta \geq 0$ and $\theta \leq 0$, that $X_n$ also satisfies this equation. But then, since we assumed that $\Lambda$ is essentially smooth, by the Gärtner-Ellis theorem (Part (c) of Thm. 2.3.6 on pg. 45 of [7]) $\{X_n\}$ satisfies an LDP with good and convex rate function the Fenchel-Legendre of $\Lambda$, namely $I$ (since we assumed $I$ to be convex).

## 10.5 A Property of Differential Entropy

In this section the logarithm is taken to the natural base for concreteness. We adopt the usual convention that $0 \log \frac{0}{0} = 0$.

### 10.5.1 The wn case

We make the general observation that for any two densities $f$ and $g$ on the real line, the negative part of the integral defining the relative entropy $K(g||f)$ is bounded above by $\frac{1}{e}$ (this is true whether or not $g$ is absolutely continuous with respect to $f$). Indeed, since we have $x \log x \geq -\frac{1}{e}$ for all $x \geq 0$, we can write

$$\int (g(x) \log \frac{g(x)}{f(x)})_- dx = \int (g(x) 1(f(x) > 0) \log \frac{g(x)}{f(x)})_- dx$$
$$= \int f(x) (\frac{g(x)}{f(x)} 1(f(x) > 0) \log \frac{g(x)}{f(x)})_- dx$$
$$\leq \int f(x) \frac{1}{e} dx \leq \frac{1}{e} .$$

Here the first step is because if $f(x) = 0$ then $g(x) \log \frac{g(x)}{f(x)}$ is either 0 or $\infty$. As, a corollary, if $S$ denotes the support of $f$, we also have

$$\int_S (g(x) \log \frac{g(x)}{f(x)})_- dx \leq \frac{1}{e} .$$

We also make the observation that for each real $x$

$$g(x) \log \frac{g(x)}{f(x)} = g(x) \log g(x) - g(x) \log f(x) . \tag{133}$$

This certainly holds if $g(x)$ and $f(x)$ are either both positive or both 0. If $g(x)$ is positive while $f(x)$ is 0 then the left hand side and the right hand side both equal $\infty$, while if $g(x)$ is 0 with $f(x)$ positive, both sides equal 0. This covers all the cases.

Let $f$ be a probability density on the real line for which

$$x \mapsto \log f(x)$$

is a bounded function on $S$, where $S$ denotes the support of $f$ (we do not need continuity of this mapping for the discussion here; continuity is only needed in order to apply the contraction principle). Let $g$ be another probability density on the real line which is absolutely continuous with respect to $f$ (i.e. $1(g(x) > 0) \subseteq 1(f(x) > 0)$). We claim that the differential entropy of $g$ is well defined.

For this, first observe that the integral

$$\int g(x) \log f(x) dx$$

is well defined. This is immediate because this integral can also be written as

$$\int_S g(x) \log f(x) dx$$

(because of the assumed absolute continuity of $g$ with respect to $f$) and since $x \mapsto \log f(x)$ is bounded as a map from $S$ to the real line, both the positive and the negative part of this integral are finite.

Using Equation (133), we may now write, for each real $x$,

$$(g(x) \log g(x))_- \leq (g(x) \log \frac{g(x)}{f(x)})_- + (g(x) \log f(x))_- \ .$$

Integrating on both sides, we conclude that $\int_S (g(x) \log g(x))_- dx < \infty$. Since

$$\int (g(x) \log g(x))_- dx = \int_S (g(x) \log g(x))_- dx \ ,$$

again by the assumed absolute continuity of $g$ with respect to $f$, the differential entropy of $g$ is well defined.

At this point we have shown that the validity of (77) in Lemma 1-wn implies the validity of (78).

More generally now, let $f$ be an arbitrary probability density on the real line and $g$ another probability density that is absolutely continuous with respect to $f$ and for which $K(g||f)$ is finite. We claim that the differential entropy of $g$ is well defined iff the integral

$$\int g(x) \log f(x) dx$$

is well defined. For one direction, note that if $\int (g(x) \log f(x))_- dx < \infty$ then Equation (133) implies that $\int (g(x) \log g(x))_- dx < \infty$, so the differential entropy of $g$ is well defined, while if $\int (g(x) \log f(x))_- dx = \infty$ and $\int g(x) \log f(x) dx$ is well defined we must have $\int (g(x) \log f(x))_+ dx < \infty$, which, with Equation (133) and the assumption that $K(g||f)$ is finite implies that $\int (g(x) \log g(x))_+ dx < \infty$ and so once again the differential entropy of $g$ is well defined. For the converse, assume that the differential entropy of $g$ is well defined. Then either $\int (-g(x) \log g(x))_+ dx$ or $\int (-g(x) \log g(x))_- dx$ must be finite. Suppose the latter is the case. From Equation (133) we can write, for each real $x$,

$$(-g(x) \log f(x))_- \leq (g(x) \log \frac{g(x)}{f(x)})_- + (-g(x) \log g(x))_- \ .$$

Integrating on both sides we conclude that $\int (-g(x) \log f(x))_- dx < \infty$, so $\int g(x) \log f(x) dx$ is well defined. In the remaining case, suppose that

$$\int (-g(x) \log g(x))_- dx = \infty,$$

while $\int(-g(x)\log g(x))_+ dx < \infty$. We also have, from Equation (133), that

$$(-g(x)\log f(x))_+ \leq (g(x)\log \frac{g(x)}{f(x)})_+ + (-g(x)\log g(x))_+ \ .$$

We now use the assumption that $K(g||f)$ is finite to see, by integrating on both sides of this equation, that $\int(-g(x)\log f(x))_+ dx < \infty$. Thus, once again $\int g(x)\log f(x)dx$ is well defined.

### 10.5.2 Markov case

The Markov version of Equation (133) is

$$b(x)q(y|x)\log \frac{q(y|x)}{p(y|x)} = b(x)q(y|x)\log q(y|x) - b(x)q(y|x)\log p(y|x) \ . \quad (134)$$

As above, we can show that if

$$(x,y) \mapsto \log p(y|x)$$

is a bounded function on $S$, where $S$ denotes the support of $a(x)p(y|x)$, and if $b(x)q(y|x)$ is a stationary Markov probability density on $\mathbb{R}^2$ which is absolutely continuous with respect to $a(x)p(y|x)$, then the differential entropy rate of $b(x)q(y|x)$, i.e. the integral

$$-\int b(x)q(y|x)\log q(y|x)dxdy$$

is well defined.

This allows us to move from the validity of (81) in Lemma 1-mn to the validity of (82).

More generally, as above we can show that if $a(x)p(y|x)$ is an arbitrary stationary Markov probability density on $\mathbb{R}^2$ and $b(x)q(y|x)$ is another stationary Markov probability density on $\mathbb{R}^2$ that is absolutely continuous with respect to $a(x)p(y|x)$ and for which the relative entropy rate $K(b(x)q(y|x)||a(x)p(y|x))$, i.e. the integral

$$\int b(x)q(y|x)\log \frac{q(y|x)}{p(y|x)}dxdy$$

is finite, then the differential entropy rate of $b(x)q(y|x)$, i.e. the integral

$$\int b(x)q(y|x)\log q(y|x)dxdy$$

is well defined iff the integral

$$\int b(x)q(y|x)\log p(y|x)dxdy$$

is well defined.

In both cases, all that is needed is to use Equation (134) everywhere that we had earlier used Equation (133) and to mimic the preceding proofs line by line.

## 10.6 Differential Entropy Counter–Example

Below, we construct an example of a pair of probability densities $g$ and $f$ such that $g$ has finite differential entropy, the relative entropy $K(f||g)$ is finite, but $f$ has undefined differential entropy.

First consider $(a_n, n \geq 3)$ and $(u_n, n \geq 3)$, sequences of nonnegative numbers given by

$$a_n = 1 , \quad u_n = \frac{1}{n(\log n)^2}$$

where the logarithm is to the natural base, for concreteness. Observe that

$$\sum_n a_n u_n < \infty ,$$

while

$$\sum_n a_n u_n \log \frac{1}{u_n} = \sum_n \frac{1}{n(\log n)^2}(\log n + 2 \log \log n) = \infty .$$

Thus, we can find a nonnegative scaling constant $\alpha > 0$ such that the piecewise constant density $f_+$ taking values $\alpha u_n$ on disjoint intervals of length $a_n$, for $n \geq 3$, integrates out to 1 and has differential entropy $\infty$. Next, consider sequences $(b_n, n \geq 3)$ and $(w_n, n \geq 3)$ given by

$$b_n = \frac{1}{n^2 (\log n)^2} , \quad w_n = n .$$

Observe that

$$\sum_n b_n w_n < \infty ,$$

while

$$\sum_n b_n w_n \log \frac{1}{w_n} = -(\sum_n \frac{1}{n^2(\log n)^2} n \log n) = -\infty .$$

Thus, we can find a nonnegative scaling constant $\gamma > 0$ such that the piecewise constant density $f_-$ taking values $\gamma w_n$ on disjoint intervals of length $b_n$, for $n \geq 3$, integrates out to 1 and has differential entropy $-\infty$. The supports of the densities $f_+$ and $f_-$ can be made disjoint, so the density $\frac{1}{2} f_+ + \frac{1}{2} f_-$ has undefined differential entropy.

We now construct a pair of probability densities $g$ and $f$ with the announced properties. First take $(a_n, n \geq 3)$ and $(u_n, n \geq 3)$ as above, and let $(v_n, n \geq 3)$ be defined by

$$v_n = \frac{1}{n(\log n)^3} .$$

Observe that

$$\sum_n a_n v_n < \infty ,$$

and

$$\sum_n a_n v_n \log \frac{1}{v_n} = \sum_n \frac{1}{n(\log n)^3}(\log n + 3 \log \log n) < \infty ,$$

and also
$$\sum_n a_n u_n \log \frac{u_n}{v_n} = \sum_n \frac{1}{n(\log n)^2} \log \log n < \infty .$$

We can therefore choose a scaling constant $\beta > 0$ such that the piecewise constant density $g_+$ taking values $\beta v_n$ on the same disjoint intervals of length $a_n$ for $n \geq 3$ as those supporting $f_+$ integrates out to 1, has finite differential entropy, and has $K(f_+||g_+) < \infty$. Next, take $(b_n, n \geq 3)$ and $(w_n, n \geq 3)$ as above, and let $(x_n, n \geq 3)$ be defined by
$$x_n = \frac{n}{\log n} .$$

Observe that
$$\sum_n b_n x_n = \sum_n \frac{1}{n^2 (\log n)^2} \frac{n}{\log n} < \infty ,$$

and
$$\sum_n b_n x_n \log \frac{1}{x_n} = \sum_n \frac{1}{n^2 (\log n)^2} \frac{n}{\log n} (\log \log n - \log n) < \infty ,$$

and
$$\sum_n b_n w_n \log \frac{w_n}{x_n} = \sum_n \frac{1}{n^2 (\log n)^2} n \log \log n < \infty .$$

We can therefore choose a scaling constant $\delta > 0$ such that the piecewise constant density $g_-$ taking values $\delta x_n$ on the same disjoint intervals of length $b_n$ for $n \geq 3$ as those supporting $f_-$ integrates out to 1, has finite differential entropy, and has $K(f_-||g_-) < \infty$. If we define $g = \frac{1}{2} g_+ + \frac{1}{2} g_-$, then $g$ is a probability density with finite differential entropy, for which $K(f||g)$ is finite when $f$ is the probability density defined earlier, but we saw already that the differential entropy of $f$ is undefined.

## 10.7 Proof of Equations (105)–(106)

If $v < \frac{\widetilde{\alpha}}{2}$, then
$$B^n \left( y^n(v), v\sigma\sqrt{n} \right) \subset B^n \left( 0, \widetilde{\alpha}\sigma\sqrt{n} \right)$$

so that $c(v) = 0$.

For $\frac{\widetilde{\alpha}}{2} < v < \frac{\widetilde{\alpha}}{\sqrt{2}}$, we have to find an upper bound on the volume of the portion of the ball of radius $v\sigma\sqrt{n}$ around the point at distance $v\sigma\sqrt{n}$ from the origin (along some ray) that is outside the ball $B^n(0, \alpha\sigma\sqrt{n})$ (this is depicted by the shaded area on Figure 1). A first upper bound on this volume is the portion of the former ball by the hyperplane perpendicular to the ray and at a distance $d\sigma\sqrt{n}$ from it (i.e. a distance of $(v+d)\sigma\sqrt{n}$ along this ray from the origin) where $d = \frac{\alpha^2}{2v} - v$ by elementary geometry. The latter portion is in turn included in a ball of radius $\sigma\sqrt{n}\sqrt{v^2 - d^2}$ (that is depicted by the dashed circle in Figure 1). Hence, $c(v) = \sqrt{v^2 - d^2}$.

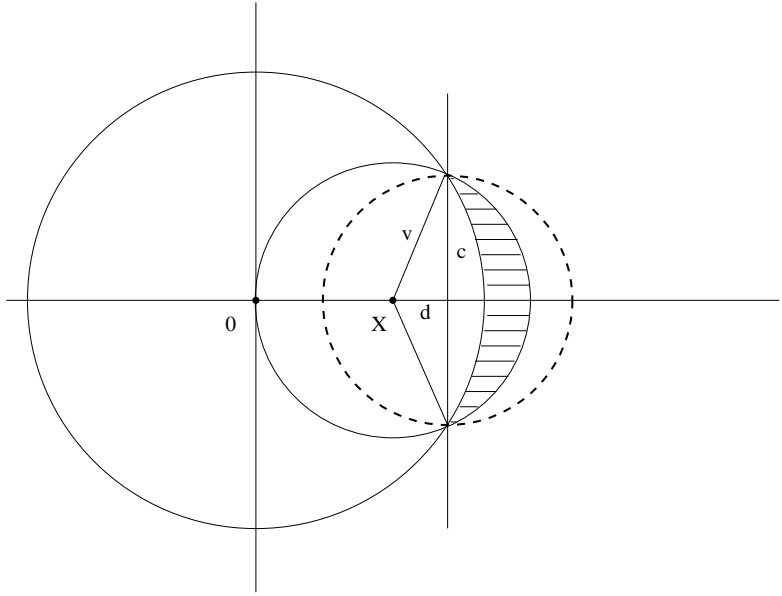

Figure 1: The origin of the plane is the tagged codeword. The large ball centered on the origin is the exclusion ball of the Matérn construction around the tagged codeword. Its radius is $(\alpha - \epsilon)\sigma\sqrt{n}$. The point $X$ is the location of the noise added to the tagged codeword. Its norm is $v\sigma\sqrt{n}$. The ball centered on $X$ with radius $v\sigma\sqrt{n}$ is the vulnerability region in the Poisson case. In the Matérn case, the vulnerability region is the shaded lune depicted on the figure. We are here in the case with $\frac{\alpha}{2} < v < \frac{\alpha}{\sqrt{2}}$. We upper bound the area of this lune by that of the ball of radius $c = \sqrt{n(v^2 - d^2)}\sigma$ with $d$ as above. This ball is depicted by the dashed line disc.

## 10.8  Proof of (112)

We have
$$\sup_{x^n : |x^n|_1 = \frac{v\sigma n}{\sqrt{2}}} W(x^n, v) \geq W(x^n(v), v)$$

with $x^n(v) = (v\sigma n/\sqrt{2}, 0, \ldots, 0)$. The region

$$R(n) = B_1^n\left(0, \frac{n\sigma(\alpha - \epsilon)}{\sqrt{2}}\right)^c \cap B_1^n\left(x^n(v), \frac{vn\sigma}{\sqrt{2}}\right)$$

includes the region

$$\left\{y^n = (y_1, \ldots, y_n) \in \mathbb{R}^n : y_1 > \frac{v\sigma n}{\sqrt{2}}, \; \frac{(\alpha n - \epsilon)\sigma n}{\sqrt{2}} < y_1 + \sum_{i=2}^n |y_i| < \frac{2v\sigma n}{\sqrt{2}}\right\}.$$

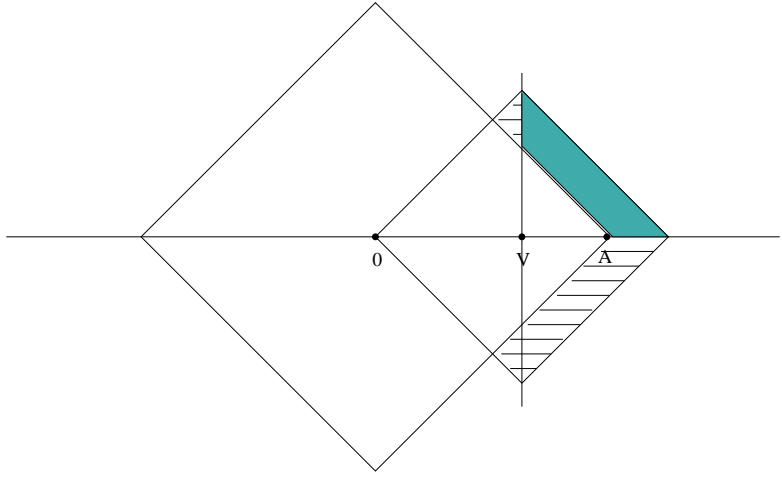

Figure 2: The origin of the plane is the tagged codeword. The large ball centered in 0 and passing through point $A$ is that with radius $\frac{n\sigma(\alpha-\epsilon)}{\sqrt{2}}$. Point $V$ is that with coordinate $x^n(v) = (v\sigma n/\sqrt{2}, 0, \ldots, 0)$. The region $R(n)$ is depicted by the union of the dashed region and the grey one. The volume $V(n)$ is that of the grey region.

This region is comprised of $2^{n-1}$ copies (one for each configuration of signs of the variables $y_2, \ldots, y_n$, see Figure 2) of the following basic region:

$$\left\{ y^n = (y_1, \ldots, y_n) \in \mathbb{R}_+^n : y_1 > \frac{v\sigma n}{\sqrt{2}},\ \frac{(\alpha-\epsilon)\sigma n}{\sqrt{2}} < y_1 + \sum_{i=2}^n y_i < \frac{2v\sigma n}{\sqrt{2}} \right\}.$$

The volume $V(n)$ of this basic region is the same as that of

$$\left\{ y^n = (y_1, \ldots, y_n) \in \mathbb{R}_+^n : \frac{(\alpha-\epsilon-v)\sigma n}{\sqrt{2}} < \sum_{i=1}^n y_i < \frac{v\sigma n}{\sqrt{2}} \right\}.$$

namely $2^{-n}$ times the volume of the $L_1$ ball of center 0 and radius $\frac{v\sigma n}{\sqrt{2}}$ deprived of the $L_1$ ball of center 0 and radius $\frac{(\alpha-\epsilon-v)\sigma n}{\sqrt{2}}$, that is

$$V(n) = 2^{-n}\left( (\sqrt{2}v\sigma)^n \frac{n^n}{n!} - (\sqrt{2}(\alpha-\epsilon-v)\sigma)^n \frac{n^n}{n!} \right).$$

Hence

$$\frac{1}{n} \ln\left( \sup_{x^n : |x^n|_1 = \frac{v\sigma n}{\sqrt{2}}} W(x^n, v) \right) \geq \frac{1}{n} \ln\left(2^{n-1} V(n)\right) \to_{n\to\infty} \ln(\sqrt{2}ve\sigma).$$

But from (111),

$$\frac{1}{n}\ln\left(\sup_{x^n:|x^n|_1=\frac{v\sigma n}{\sqrt{2}}} W(x^n,v)\right) \leq \frac{1}{n}\ln\left(\mathrm{Vol}B_1^n(0,\frac{v\sigma n}{\sqrt{2}})\right) \to_{n\to\infty} \ln(\sqrt{2}ve\sigma).$$

This completes the proof of (112).

## Acknowledgment

The work of the first author was supported by NSF grants CCF-0500234, CNS-0627161 and CCF-0635372, by Marvell Semiconductor, and by the University of California MICRO program during the initial stages of this project, and is current supported by the ARO MURI grant W911NF-08-1-0233, "Tools for the Analysis and Design of Complex Multi-Scale Networks", by the NSF grant CNS-0910702, by the NSF Science & Technology Center grant CCF-0939370, "Science of Information", by Marvell Semiconductor Inc., and by the U. C. Discovery program.

This work started when the second author was a Visiting Miller Professor at UC Berkeley.

Both authors would like to thank the Isaac Newton Institute for providing a wonderful environment during Spring 2010 as part of the program "Stochastic Processes in Communications Sciences", where much of the final part of this work was done.